\documentclass[11pt,fleqn]{article}
\usepackage{amssymb}
\usepackage{graphicx}
\usepackage{amsmath}
\usepackage{times}

\input{tcilatex}

\begin{document}

\title{Efficient FMM accelerated vortex methods in three dimensions \\
via the Lamb-Helmholtz decomposition}
\author{Nail A. Gumerov\footnote{%
Corresponding Author. Also at Fantalgo, LLC., Elkridge, MD 21075; E-mail: 
\texttt{gumerov@umiacs.umd.edu}; Phone: +1-301-405-8210; Fax:
+1-301-314-9658; web: \texttt{http://www.umiacs.umd.edu/users/gumerov}} \ \
and \ Ramani Duraiswami\footnote{%
Also Department of Computer Science, and at Fantalgo, LLC. E-mail:\texttt{%
ramani@umiacs.umd.edu}, web: \texttt{http://www.umiacs.umd.edu/users/ramani}}%
\\
Institute for Advanced Computer Studies \\
University of Maryland\\
College Park, MD 20742, USA}
\maketitle

\begin{abstract}
Vortex methods are used to efficiently simulate incompressible
flows using Lagrangian techniques. Use of the FMM (Fast Multipole Method)
allows considerable speed up of both velocity evaluation and vorticity
evolution terms in these methods. Both equations require field evaluation of
constrained (divergence free) vector valued quantities (velocity, vorticity)
and cross terms from these. These are usually evaluated by performing
several FMM accelerated sums of scalar harmonic functions.

We present a formulation of vortex methods based on the Lamb-Helmholtz
decomposition of the velocity in terms of two scalar potentials. In its
original form, this decomposition is not invariant with respect to
translation, violating a key requirement for the FMM. One of the key
contributions of this paper is a theory for translation for this
representation. The translation theory is developed by introducing
``conversion'' operators, which enable the representation to be restored in
an arbitrary reference frame. Using this form, efficient vortex element
computations can be made, which need evaluation of just two scalar harmonic
FMM sums for evaluating the velocity and vorticity evolution terms. Details
of the decomposition, translation and conversion formulae, and sample
numerical results are presented.
\end{abstract}

\section{Introduction}

Vortex methods are used to simulate the Navier Stokes equation in the velocity--vorticity form with Lagrangian discretization. Since vortex
particles are initially placed only in the region of finite vorticity and
can convect along with the flow, these methods provide an optimized spatial
discretization. Consider an incompressible flow generated by a set of $N$
vortex elements, characterized by coordinates of the centers (sources) $%
\mathbf{x}_{i}$ and constant strength vector $\boldsymbol{\omega }_{i},$ $%
i=1,...,N.$ Each element centered at location $\mathbf{x}_{i}$ produces an
elementary velocity field $\mathbf{v}_{i}\left( \mathbf{y}\right)$ according
to the Biot-Savart law, and the total velocity field can be computed as a
superposition of such elementary fields (e.g., see \cite{Batchelor1967}): 
\begin{equation}
\mathbf{v}\left( \mathbf{y}\right) =\sum_{i=1}^{N}\mathbf{v}_{i}\left( 
\mathbf{y}\right) ,\quad \mathbf{v}_{i}\left( \mathbf{y}\right) =\frac{%
\boldsymbol{\omega }_{i}\times \left( \mathbf{y} - \mathbf{x}_{i} \right) }{%
\left\vert \mathbf{y} - \mathbf{x}_{i} \right\vert ^{3}}=\nabla \times \frac{%
\boldsymbol{\omega }_{i}}{\left\vert \mathbf{y}-\mathbf{x}_{i}\right\vert }.
\label{1}
\end{equation}
In practice this field needs to evaluated at $M$ evaluation points, $\mathbf{%
y}_{j}$, which has $O(MN)$ cost. The Biot-Savart kernel is composed of a
vector of dipole solutions of the Laplace equation. It is well known that
the Fast Multipole Method (FMM) can be used to evaluate such sums to any
specified accuracy $\epsilon $ at a $O(N+M)$ reduced cost \cite%
{Greengard1987:JCP}.

The vortex elements move with the flow. This motion also causes an evolution
of the vortex field according to the vortex evolution equation. For inviscid
flow, the evolution equations for the vortex positions and strengths
respectively are 
\begin{eqnarray}
\frac{d\mathbf{x}_{i}}{dt} &=&\left. \mathbf{v}\right|_{\mathbf{x}=
\mathbf{x}_{i}},\quad \quad \frac{d\boldsymbol{\omega }_{i}}{dt}=\mathbf{s}%
_{i},\quad  \label{s1.1} \\
\mathbf{s}_{i} &=&\boldsymbol{\omega }_{i}\cdot \left. \nabla \mathbf{v}%
\right| _{\mathbf{x}=\mathbf{x}_{i}}\mathbf{,\quad v}\left( \mathbf{x}_{i}%
\mathbf{;}t\right) =\sum_{j\neq i}\mathbf{v}_{j}\left( \mathbf{x}%
_{i};t\right) .  \notag
\end{eqnarray}
Here the right hand side for the vortex strength is the so-called vortex
stretching term, the evaluation of which requires the
gradient of the velocity vector. The evolution equation for the vorticity
can be modified to account for liquid viscosity. Also the elementary
velocity field in Eq. (\ref{1}) can be modified using a smoothing kernel, $%
K(\left| \mathbf{y-x}_{i}\right| ;a)$, 
\begin{equation}
\mathbf{v}_{i}\left( \mathbf{y;}a\right) =\frac{\boldsymbol{\omega}%
_{i}\times \left( \mathbf{y}-\mathbf{x}_{i}\right) }{\left| \mathbf{y}-%
\mathbf{x}_{i}\right| ^{3}}K(\left| \mathbf{y-x}_{i}\right| ;a),\quad
K(r;a)=1+O(\epsilon ),\quad r\gg a,  \label{2}
\end{equation}
which has an effect only on the near field, $r\lesssim a,$ where $a$ is the
radius of the vortex core, which may change in time, and $\epsilon \ll 1$ is
the tolerance for approximation. This modification does not affect the far
field, which can also be computed with the tolerance $O\left( \epsilon \right) $ in an accelerated manner via the FMM. In the
sequel we do not specify the core function $K(r;a)$; several choices
including the Gaussian and polynomial forms are discussed in the literature
(see e.g., \cite{Cottet2000:Book, Koumoutsakos2005:ARFM,
Winckelmans2005:CRP, Yokota2009:CPC}), and there are several ways to speed
up the local summation as well. Extensions to compressible flow are also
possible \cite{Eldredge2002:JCP}.

The evolution equation is integrated using an appropriate time stepping
scheme. The right hand side of this equation, also results in an $N$ body
computation for the influence of particles in the far-field, with somewhat
more complicated terms. As discussed above, the FMM for the vortex element
method is closely related to the scalar FMM\ for sums of multipoles of the
Laplace equation (\textquotedblleft harmonic FMM\textquotedblright ). In
fact, it is possible to start with a program for a
harmonic FMM and appropriately modify it to create a fast vortex method. In terms of performance, computation of potential gradients and
higher derivatives can be referred as \textquotedblleft
auxiliary\textquotedblright\ computations, which can be done as soon as
local expansions for the potentials are available. In this sense the question as to
how many independent potentials need to be computed to obtain right hand
sides of the evolution equations (\ref{s1.1}) is important. For example, to compute the three components of the gravitational force in an stellar $N$-body computation one needs only one harmonic FMM; the gradients can be obtained via differentiation of expansions, which is done efficiently by application of sparse matrix operators to the potential expansions. Treating the problem in a straightforward way, for the vortex method, one
should use three independent harmonic FMM sums (for each velocity
component). However, because of the divergence constraint, one may speculate that it is possible to reduce this number to two. THat this in fact is so is a main result of the present paper. 

A similar reduction of the complexity
to solution of two harmonic FMMs was obtained for the biharmonic equation 
\cite{Gumerov2006:JCP} (opposed to five FMMs using factorization \cite%
{Fu2000:CNME}). For the Stokes equations, where the solution can be
decomposed to the sum of Stokeslets and Stresslets, a representation via
three harmonic potentials (Lamb-Helmholtz decomposition \cite{Lamb1932})
requires only three harmonic FMMs (see also \cite{Sangani1996}), while a
more simple way based on factorization \cite{Tornberg2008:JCP} shows that
the evaluation can be done with a cost of four harmonic FMM calls. 

In this paper we develop such an efficient version of the FMM for vortex methods, which achieves an evaluation of both the velocity and
stretching term sums at a cost of only two scalar harmonic FMMs (this also
can be reduced to one complex valued harmonic FMM since the physical fields
are real). Our approach is based on the Lamb-Helmholtz decomposition \cite%
{Lamb1932}, which allows representation of the vector field in the form of
two scalar potential fields. This form is however not invariant to
translation, and cannot be used as is, with an FMM
summation algorithm. We develop conversion operators that allow this form to
be translated.

Section 2 of the paper introduces the problem and notation, and shows that
the equations can be considered to be solutions of a \emph{divergence
constrained vector Laplace equation}. Section 3 develops the translation
theory for such equations, which is the main mathematical result of the
paper. Section 4 shows how the new translation theory can be used, together
with a harmonic FMM, to create an  FMM for vortex methods. Section 5
presents the results of numerical testing and some examples of FMM
accelerated vortex element method computations. Section 6 concludes the
paper. Mathematical details are provided in appendices.

\section{Statement of the problem}

We are given $N$ vortex blobs of strength $\boldsymbol{\omega }_{i},$ $%
i=1,...,N$ located at points $\mathbf{x}_{i}$ and moving with the flow. The
velocity field can be evaluated using either Eq. (\ref{1}) or (\ref{2}),
which both have the same asymptotic far-field form. The evolution of the
vortex positions and the vortex strengths is given by Eqs. (\ref{s1.1}). At $%
\mathbf{y\neq x}_{i},$ $i=1,...,N$, the velocity field $\mathbf{v}\left( 
\mathbf{y}\right) $ satisfies the divergence constrained vector Laplace
equation (DCVLE) 
\begin{equation}
\nabla ^{2}\mathbf{v=0,\quad }\nabla \cdot \mathbf{v}=0.  \label{3}
\end{equation}
If the divergence of the field were not constrained, each Cartesian
component of the velocity would be an independent harmonic function. The
divergence constraint, however, reduces the degree of freedom for solutions
by 1, and, in fact, only two harmonic scalar potentials, $\phi $ and $\chi $%
, are necessary to describe the total field inside or outside a sphere
centered at the origin 
\begin{equation}
\mathbf{v}\left( \mathbf{r}\right) =\nabla \phi \left( \mathbf{r}\right)
+\nabla \times \left( \mathbf{r}\chi \left( \mathbf{r}\right) \right) ,\quad
\nabla ^{2}\phi =0,\quad \nabla ^{2}\chi =0.  \label{4}
\end{equation}
This decomposition can be treated as a general Helmholtz decomposition of an
arbitrary vector field. Presumably, this form is due to Lamb \cite{Lamb1932}%
, who used it to obtain a general solution for the Stokes equations in
spherical coordinates, and we refer to this as the Lamb-Helmholtz
decomposition. Indeed, Eq. (\ref{3}) are the Stokes equations with zero
pressure for which the Lamb solution provides (\ref{4}).

The DCVLE appears naturally when one attempts to follow a general procedure to reconstruct an arbitrary vector field from given curl, $%
\boldsymbol{\omega}\left(\mathbf{r}\right) $, and divergence, $q\left( 
\mathbf{r}\right) $, 
\begin{equation}
\nabla \times \mathbf{v=\omega ,\quad }\nabla \cdot \mathbf{v}=q.  \label{5}
\end{equation}
The solution of these equations in free space can be written in the
form (e.g. see \cite{Batchelor1967}): 
\begin{equation}
\mathbf{v}\left( \mathbf{y}\right) =-\nabla _{\mathbf{y}} \int_{V}\frac{%
q\left( \mathbf{x}\right) }{4\pi \left| \mathbf{y}-\mathbf{x}\right| }%
dV\left( \mathbf{x}\right) +\nabla _{\mathbf{y}}\times \int_{V}\frac{%
\boldsymbol{\omega }\left( \mathbf{x}\right) }{4\pi \left| \mathbf{y}- 
\mathbf{x}\right| }dV\left( \mathbf{x}\right) .  \label{6}
\end{equation}
Subdividing the space to the vicinity of evaluation point $\mathbf{y}$ (near
field) and the domain outside this neighborhood (far field) and discretizing
the integrals for the far field using quadratures with weights $w_{i}$ and
nodes $\mathbf{x}_{i}$, we obtain for the far field contribution 
\begin{eqnarray}
\mathbf{v}_{f}\left( \mathbf{y}\right) &=&\sum_{i}\mathbf{v}_{fi}\left( 
\mathbf{y}\right) ,\quad \mathbf{v}_{fi}\left( \mathbf{y}\right) =-\nabla 
\frac{q_{i}}{\left| \mathbf{y}-\mathbf{x}_{i}\right| }+\nabla \times \frac{%
\boldsymbol{\omega }_{i}}{\left| \mathbf{y}-\mathbf{x}_{i}\right| },
\label{7} \\
q_{i} &=&\frac{w_{i}q\left( \mathbf{x}_{i}\right) }{4\pi },\quad \mathbf{%
\omega }_{i}=\frac{w_{i}\boldsymbol{\omega }\left( \mathbf{x}_{i}\right) }{%
4\pi }.  \notag
\end{eqnarray}
Hence, the far field satisfies Eq. (\ref{3}) for which decomposition (\ref{4}%
) can be used and just an addition to potential $\phi $ due to a given
monopole distribution $q\left( \mathbf{x}\right) $ provides solution for a
general case. As mentioned, in the present paper we do not address
computation of the near field, which can be done locally, e.g. using
appropriate smoothing kernels. Note that solution (\ref{6}) of Eq. (\ref{5})
is unique up to a gradient of a harmonic function $\Phi $, which should be
found from the boundary conditions. Such functions for a given boundary can
be added to $\phi \left( \mathbf{r}\right) $ in Eq. (\ref{4}).

Equations (\ref{5}) with $q\neq 0$ appear, e.g. in vortex methods for
compressible flows. An example of equations (for 2D) can be found in \cite%
{Eldredge2002:JCP}, which can be appropriately modified for 3D. In terms of
computational complexity, besides the velocity field and stretching term
computations, also contraction of the velocity gradient tensor, $\beta =$ $%
\nabla \mathbf{v:}\nabla \mathbf{v}$ should be computed in this case. This
term can be computed simultaneously with computation of the vortex
stretching term. Thus the cost in these extended cases should remain the
same.

\section{Translation theory for DCVLE}

\subsection{Basic translation and differential operators}

\textbf{Translation operator:}\ A generic translation or shift operator $%
\mathcal{T}(\mathbf{t})$, where $\mathbf{t}$ is a constant termed the
translation vector, acts on some scalar valued function $\phi \left( \mathbf{%
r}\right) ,$ to produce a new function $\widehat{\phi }\left( \mathbf{r}%
\right) $ (the translate), whose values coincide with $\phi \left( \mathbf{r}%
\right) $ at shifted values of the argument 
\begin{equation}
\widehat{\phi }\left( \mathbf{r}\right) =\mathcal{T}(\mathbf{t})\left[ \phi
\left( \mathbf{r}\right) \right] =\mathcal{T}_{\mathbf{t}}\phi ,\quad 
\widehat{\phi }\left( \mathbf{r}\right) =\phi \left( \mathbf{r+t}\right)
,\quad \mathbf{r,t\in }\mathbb{R}^{3}.  \label{bas1}
\end{equation}
This operator is linear. Also the translates of harmonic functions are also
harmonic functions.

\textbf{Elementary directional differential operators:} We introduce the
following notation for differential operators which appear in derivations: 
\begin{equation}
\mathcal{D}_{\mathbf{r}}=\mathbf{r}\cdot \nabla ,\quad \mathcal{D}_{\mathbf{t%
}}=\mathbf{t}\cdot \nabla ,\quad \mathcal{D}_{\mathbf{r\times t}}=\left( 
\mathbf{r\times t}\right) \cdot \nabla .  \label{bas2}
\end{equation}
It can be shown that if $\phi \left( \mathbf{r}\right) $ is a harmonic
function in some domain, then $\mathcal{D}_{\mathbf{r}}\phi $ , $\mathcal{D}%
_{\mathbf{t}}\phi $, and $\mathcal{D}_{\mathbf{r\times t}}\phi $ are also
harmonic functions in the same domain. Note also that operators $\mathcal{D}%
_{\mathbf{t}}$ and $\mathcal{D}_{\mathbf{r\times t}}$ are related to an
infinitesimal translation in the direction of vector $\mathbf{t}$ and an
infinitesimal rotation about axis $\mathbf{t}$.

\subsection{Conversion operators for the DCVLE}

Consider the translation of the vector $\mathbf{v}\left( \mathbf{r}%
\right) $ in (\ref{4}). Using (\ref{bas1}) for the
translated functions,  
\begin{align}
\widehat{\mathbf{v}}\left( \mathbf{r}\right) & =\mathbf{v}\left( \mathbf{r+t}%
\right) =\nabla \phi \left( \mathbf{r+t}\right) +\nabla \times \left( \left( 
\mathbf{r+t}\right) \chi \left( \mathbf{r+t}\right) \right)  \label{CVL1} \\
& =\nabla \widehat{\phi }\left( \mathbf{r}\right) +\nabla \times \left(
\left( \mathbf{r+t}\right) \widehat{\chi }\left( \mathbf{r}\right) \right)
=\nabla \widehat{\phi }\left( \mathbf{r}\right) +\nabla \times \left( 
\mathbf{r}\widehat{\chi }\left( \mathbf{r}\right) \right) +\nabla \times
\left( \mathbf{t}\widehat{\chi }\left( \mathbf{r}\right) \right) .  \notag
\end{align}
Obviously this is not form (\ref{4}) representing $\widehat{\mathbf{v}}%
\left( \mathbf{r}\right) $. Our goal is to find harmonic functions $%
\widetilde{\phi }$ and $\widetilde{\chi }$, which provide such a
representation, i.e. 
\begin{equation}
\widehat{\mathbf{v}}\left( \mathbf{r}\right) =\nabla \widetilde{\phi }%
+\nabla \times \left( \mathbf{r}\widetilde{\chi }\right) .  \label{CVL2}
\end{equation}
For this purpose we introduce ``conversion'' operators $\mathcal{C}_{ij},$ $%
i,j=1,2:$ 
\begin{equation}
\widetilde{\phi }=\mathcal{C}_{11}\widehat{\phi }+\mathcal{C}_{12}\widehat{%
\chi },\quad \widetilde{\chi }=\mathcal{C}_{21}\widehat{\phi }+\mathcal{C}%
_{22}\widehat{\chi },  \label{CVL3}
\end{equation}
which are linear due to the linearity of all transforms considered.

Comparing representations (\ref{CVL1}) and (\ref{CVL2}) we deduce, that $%
\widehat{\phi }\left( \mathbf{r}\right) $ contributes only to $\widetilde{%
\phi }\left( \mathbf{r}\right) ,$ leading to 
\begin{equation}
\mathcal{C}_{11}=\mathcal{I},\quad \mathcal{C}_{21}=0,  \label{CVL4}
\end{equation}
where $\mathcal{I}$ is the identity operator. So, we can introduce harmonic
functions $\phi ^{\prime }$ and $\chi ^{\prime }$ according to the following
relations 
\begin{align}
\widetilde{\phi }& =\widehat{\phi }+\mathcal{C}_{12}\widehat{\chi }=\widehat{%
\phi }+\phi ^{\prime },  \label{CVL5} \\
\widetilde{\chi }& =\mathcal{C}_{22}\widehat{\chi }=\widehat{\chi }+\chi
^{\prime }.  \notag
\end{align}
Having two representations of $\widehat{\mathbf{v}}$, (\ref{CVL1}) and (\ref%
{CVL2}), and using Eq. (\ref{CVL5}), we obtain 
\begin{equation}
\nabla \phi ^{\prime }+\nabla \times \left( \mathbf{r}\chi ^{\prime }\right)
=\nabla \times \left( \mathbf{t}\widehat{\chi }\right) .  \label{CVL6}
\end{equation}
Taking scalar product with $\mathbf{r}$ and noticing that $\mathbf{r}\cdot
\nabla \times \left( \mathbf{r}\chi ^{\prime }\right) =0$, one can see that 
\begin{equation}
\mathbf{r}\cdot \nabla \phi ^{\prime }=\mathbf{r}\cdot \nabla \times \left( 
\mathbf{t}\widehat{\chi }\right) =\mathbf{r}\cdot \left( \nabla \widehat{%
\chi }\times \mathbf{t}\right) =-\left( \mathbf{r\times t}\right) \cdot
\nabla \widehat{\chi }.  \label{CVL7}
\end{equation}
Another relation can be obtained if we take the curl of expression (\ref%
{CVL6}): 
\begin{equation}
\nabla \times \nabla \times \left( \mathbf{r}\chi ^{\prime }\right) =\nabla
\times \nabla \times \left( \mathbf{t}\widehat{\chi }\right) .  \label{CVL8}
\end{equation}
It is not difficult to check that the following identities hold for the
harmonic functions $\chi ^{\prime }$ and $\widehat{\chi }$: 
\begin{eqnarray}
\nabla \times \nabla \times \left( \mathbf{r}\chi ^{\prime }\right)
&=&\nabla \left( \chi ^{\prime }+\mathbf{r}\cdot \nabla \chi ^{\prime
}\right) ,  \label{CVL9} \\
\nabla \times \nabla \times \left( \mathbf{t}\widehat{\chi }\right)
&=&\nabla \left( \mathbf{t}\cdot \nabla \widehat{\chi }\right) .  \notag
\end{eqnarray}
Note that all scalar potentials are defined up to a constant.
Therefore, we obtain from Eqs (\ref{CVL8}) and (\ref{CVL9}): 
\begin{equation}
\chi ^{\prime }+\mathbf{r}\cdot \nabla \chi ^{\prime }=\mathbf{t}\cdot
\nabla \widehat{\chi }.  \label{CVL10}
\end{equation}
Using (\ref{bas2}) we can rewrite relations (\ref{CVL7}) and (\ref{CVL10})
in the form 
\begin{equation}
\mathcal{D}_{\mathbf{r}}\phi ^{\prime }=-\mathcal{D}_{\mathbf{r\times t}}%
\widehat{\chi },\quad \left( \mathcal{I}+\mathcal{D}_{\mathbf{r}}\right)
\chi ^{\prime }=\mathcal{D}_{\mathbf{t}}\widehat{\chi }.  \label{CVL11}
\end{equation}
In the next sections, we show that operators $\mathcal{D}_{\mathbf{r}}$ and $%
\left( \mathcal{I}+\mathcal{D}_{\mathbf{r}}\right) $ are invertible, and so we can write 
\begin{eqnarray}
\widetilde{\phi } &=&\widehat{\phi }-\mathcal{D}_{\mathbf{r}}^{-1}\mathcal{D}%
_{\mathbf{r\times t}}\widehat{\chi },  \label{CVL13} \\
\widetilde{\chi } &=&\widehat{\chi }+\left( \mathcal{I}+\mathcal{D}_{\mathbf{%
r}}\right) ^{-1}\mathcal{D}_{\mathbf{t}}\widehat{\chi }.  \notag
\end{eqnarray}
Comparing Eqs (\ref{CVL5}) and (\ref{CVL13}), we obtain the following
expressions for the conversion operators 
\begin{equation}
\mathcal{C}_{12}\left( \mathbf{t}\right) =-\mathcal{D}_{\mathbf{r}}^{-1}%
\mathcal{D}_{\mathbf{r\times t}},\quad \mathcal{C}_{22}\left( \mathbf{t}%
\right) =\mathcal{I}+\left( \mathcal{I}+\mathcal{D}_{\mathbf{r}}\right) ^{-1}%
\mathcal{D}_{\mathbf{t}}.  \label{CVL14}
\end{equation}

\subsection{Expansions of harmonic functions}

In addition to Cartesian coordinates we will use spherical coordinates $%
\left( r,\theta ,\varphi \right) $: 
\begin{equation}
\mathbf{r}=\mathbf{(}x,y,z)=r\left( \sin \theta \cos \varphi ,\sin \theta
\sin \varphi ,\cos \theta \right) ,\quad \mathbf{r\in }\mathbb{R}^{3}.
\label{rsbf1}
\end{equation}
For expansions of the solutions of the Laplace equation that are regular
inside or outside a sphere centered at the origin of the reference frame we
introduce the regular or local functions $R_{n}^{m}\left( \mathbf{r}\right) $
and singular or multipole functions $S_{n}^{m}\left( \mathbf{r}\right) $,
that are respectively defined as 
\begin{align}
R_{n}^{m}\left( \mathbf{r}\right) & =\frac{\left( -1\right) ^{n}i^{\left|
m\right| }}{(n+\left| m\right| )!}r^{n}P_{n}^{\left| m\right| }(\mu
)e^{im\varphi },\quad \mu =\cos \theta ,  \label{rsbf2} \\
S_{n}^{m}\left( \mathbf{r}\right) & =\frac{i^{-\left| m\right| }(n-\left|
m\right| )!}{r^{n+1}}P_{n}^{\left| m\right| }(\mu )e^{im\varphi },\quad
n=0,1,...,\quad m=-n,...,n,  \notag
\end{align}
where $P_{n}^{m}(\mu )$ are the associated Legendre functions, defined by
Rodrigues' formula 
\begin{equation}
P_{n}^{m}\left( \mu \right) =\frac{\left( -1\right) ^{m}\left( 1-\mu
^{2}\right) ^{m/2}}{2^{n}n!}\frac{d^{m+n}}{d\mu ^{m+n}}\left( \mu
^{2}-1\right) ^{n},\text{ }m\geqslant 0.  \label{rsbf3}
\end{equation}
These functions are related to each other via 
\begin{equation}
S_{n}^{m}\left( \mathbf{r}\right) =\left( -1\right) ^{n+m}\left( n-m\right)
!(n+m)!r^{-2n-1}R_{n}^{m}\left( \mathbf{r}\right) .  \label{rsbf4}
\end{equation}

The functions $R_{n}^{m}\left( \mathbf{r}\right) $ and $S_{n}^{m}\left( 
\mathbf{r}\right) $ defined above coincide with the normalized basis
functions $I_{n}^{m}\left( \mathbf{r}\right) $ and $O_{n}^{m}\left( \mathbf{r%
}\right) $ considered in \cite{Epton1995:SISC} and normalized spherical
basis functions in \cite{Gumerov2006:JCP}. The expansion of the Green's
function in this basis is 
\begin{equation}
\left| \mathbf{r}-\mathbf{r}_{0}\right| ^{-1}=\sum_{n=0}^{\infty
}\sum_{m=-n}^{n}R_{n}^{-m}(-\mathbf{r}_{0})S_{n}^{m}(\mathbf{r}),\quad
r>r_{0}.  \label{rsbf5}
\end{equation}
Further, we represent harmonic functions in terms of sets of expansion
coefficients over a certain basis centered at a given point, e.g. the local
and multipole expansions centered at the origin are 
\begin{equation}
\phi \left( \mathbf{r}\right) =\sum_{n=0}^{\infty }\sum_{m=-n}^{n}\phi
_{n}^{m}F_{n}^{m}(\mathbf{r}),\quad \chi \left( \mathbf{r}\right)
=\sum_{n=0}^{\infty }\sum_{m=-n}^{n}\chi _{n}^{m}F_{n}^{m}(\mathbf{r}),\quad
F=R,S.  \label{rsbf6}
\end{equation}
Absolute and uniform convergence of these series in the expansion regions is
assumed below. We also extend the definition of the basis functions for
arbitrary order $m,$ to shorten some expressions 
\begin{equation}
R_{n}^{m}\left( \mathbf{r}\right) =S_{n}^{m}\left( \mathbf{r}\right)
=0,\quad \left| m\right| >n,\quad n=0,1,...  \label{rsbf7}
\end{equation}

\subsection{Matrix representation of operators}

Let $\mathcal{L}$ be a linear operator, such that for harmonic function $%
\phi $, $\psi =\mathcal{L}\phi $ is also a harmonic function. Assume further
that both $\phi $ and $\psi $ can be expanded into series of type (\ref%
{rsbf6}). There should be a linear relation between the expansion
coefficients $\mathbf{\Psi =}\left\{ \psi _{n}^{m}\right\} $ and $\mathbf{%
\Phi }=\left\{ \phi _{n}^{m}\right\} $, which, generally speaking, will have
a form $\mathbf{\Psi =L\Phi }$, where $\mathbf{L}$ is a matrix, or
representation of $\mathcal{L}$. Of course, for a given $\mathcal{L}$ the
matrix $\mathbf{L}$ depends on the bases over which the expansion is taken.

Let $\phi \left( \mathbf{r}\right) $ be expanded over basis $\left\{
F_{n}^{m}(\mathbf{r})\right\} $, while $\psi \left( \mathbf{r}\right) $ be
expanded over basis $\left\{ G_{n}^{m}(\mathbf{r})\right\} $. The action of
the operator $\mathcal{L}$ on a basis function $F_{n}^{m}(\mathbf{r})$ can
be represented as 
\begin{equation}
\mathcal{L}F_{n}^{m}(\mathbf{r})=\sum_{n^{\prime }=0}^{\infty
}\sum_{m^{\prime }=-n^{\prime }}^{n^{\prime }}L_{n^{\prime }n}^{m^{\prime
}m}G_{n^{\prime }}^{m^{\prime }}\left( \mathbf{r}\right) ,\quad
n=0,1,...,\quad m=-n,...,n,  \label{mr1}
\end{equation}
where $L_{n^{\prime }n}^{m^{\prime }m}$ are the reexpansion coefficients.
It can be shown that the entries of matrix $\mathbf{L}$
are $L_{nn^{\prime }}^{mm^{\prime }}$, i.e. $\mathbf{L}$ is the matrix
transpose of the matrix of reexpansion coefficients. Indeed, 
\begin{eqnarray}
\sum_{n=0}^{\infty }\sum_{m=-n}^{n}\psi _{n}^{m}G_{n}^{m}(\mathbf{r})
&=&\psi \left( \mathbf{r}\right) =\mathcal{L}\phi \left( \mathbf{r}\right)
=\sum_{n^{\prime }=0}^{\infty }\sum_{m^{\prime }=-n^{\prime }}^{n^{\prime
}}\phi _{n^{\prime }}^{m^{\prime }}\mathcal{L}F_{n^{\prime }}^{m^{\prime }}(%
\mathbf{r})  \label{mr2} \\
&=&\sum_{n^{\prime }=0}^{\infty }\sum_{m^{\prime }=-n^{\prime }}^{n^{\prime
}}\phi _{n^{\prime }}^{m^{\prime }}\sum_{n=0}^{\infty
}\sum_{m=-n}^{n}L_{nn^{\prime }}^{mm^{\prime }}G_{n}^{m}\left( \mathbf{r}%
\right) =\sum_{n=0}^{\infty }\sum_{m=-n}^{n}\left[ \sum_{n^{\prime
}=0}^{\infty }\sum_{m^{\prime }=-n^{\prime }}^{n^{\prime }}L_{nn^{\prime
}}^{mm^{\prime }}\phi _{n^{\prime }}^{m^{\prime }}\right] G_{n}^{m}\left( 
\mathbf{r}\right) .  \notag
\end{eqnarray}
The reexpansion coefficients for the translation operator $\mathcal{T}_{%
\mathbf{t}}$, Eq. (\ref{bas1}), in the local and multipole bases (\ref{rsbf2}%
) can be simply expressed via the respective basis functions (see \cite%
{Epton1995:SISC} and \cite{Gumerov2006:JCP}): 
\begin{eqnarray}
\mathcal{T}_{\mathbf{t}}F_{n}^{m}(\mathbf{r}) &=&\sum_{n^{\prime
}=0}^{\infty }\sum_{m^{\prime }=-n^{\prime }}^{n^{\prime }}\left( F|G\right)
_{n^{\prime }n}^{m^{\prime }m}\left( \mathbf{t}\right) G_{n^{\prime
}}^{m^{\prime }}\left( \mathbf{r}\right) ,\quad F,G=S,R,  \label{mr3} \\
\left( R|R\right) _{n^{\prime }n}^{m^{\prime }m}\left( \mathbf{t}\right)
&=&R_{n-n^{\prime }}^{m-m^{\prime }}\left( \mathbf{t}\right) ,\quad \left(
S|R\right) _{n^{\prime }n}^{m^{\prime }m}\left( \mathbf{t}\right)
=S_{n+n^{\prime }}^{m-m^{\prime }}\left( \mathbf{t}\right) ,\quad \left(
S|S\right) _{n^{\prime }n}^{m^{\prime }m}\left( \mathbf{t}\right)
=R_{n^{\prime }-n}^{m-m^{\prime }}\left( \mathbf{t}\right) .  \notag
\end{eqnarray}
Here $\left( R|R\right) _{n^{\prime }n}^{m^{\prime }m}$, $\left( S|R\right)
_{n^{\prime }n}^{m^{\prime }m}$, and $\left( S|S\right) _{n^{\prime
}n}^{m^{\prime }m}$ are the entries of the local-to-local (L2L),
multipole-to-local (M2L), and multipole-to-multipole (M2M) translation
matrices, respectively. To obtain representations of other operators
appeared above, we use differential relations for the basis functions, which
also can be found in \cite{Epton1995:SISC} and \cite{Gumerov2006:JCP}: 
\begin{eqnarray}
\mathcal{D}_{z}R_{n}^{m}\left( \mathbf{r}\right) &=&-R_{n-1}^{m}\left( 
\mathbf{r}\right) ,\quad \mathcal{D}_{z}S_{n}^{m}\left( \mathbf{r}\right)
=-S_{n+1}^{m}\left( \mathbf{r}\right) ,  \label{mr4} \\
\mathcal{D}_{x+iy}R_{n}^{m}\left( \mathbf{r}\right) &=&iR_{n-1}^{m+1}\left( 
\mathbf{r}\right) ,\quad \mathcal{D}_{x+iy}S_{n}^{m}\left( \mathbf{r}\right)
=iS_{n+1}^{m+1}\left( \mathbf{r}\right) ,  \notag \\
\mathcal{D}_{x-iy}R_{n}^{m}\left( \mathbf{r}\right) &=&iR_{n-1}^{m-1}\left( 
\mathbf{r}\right) ,\quad \mathcal{D}_{x-iy}S_{n}^{m}\left( \mathbf{r}\right)
=iS_{n+1}^{m-1}\left( \mathbf{r}\right) .  \notag
\end{eqnarray}
where 
\begin{equation}
\mathcal{D}_{x\pm iy}=\frac{\partial }{\partial x}\pm i\frac{\partial }{%
\partial y},\quad \mathcal{D}_{z}=\frac{\partial }{\partial z}.  \label{mr5}
\end{equation}
Appendix A provides explicit matrix representation of operators (\ref{bas2})
and (\ref{CVL14}) required for numerical implementation of the present
method.

\section{Fast multipole method}

There is an extensive literature on the FMM for the 3D Laplace equation
(see, e.g., \cite{Greengard1987:JCP, Cheng1999:JCP, Greengard1997:Acta,
Gumerov2005:TR, Gumerov2008:JCP}), and we just present the
modifications necessary to use this harmonic FMM for
vortex methods.

Note that the FMM can be considered as a way to perform a dense matrix-vector product based
on decomposition of the matrix into sparse and dense parts 
\begin{eqnarray}
\mathbf{v(y}_{j}) &=&\sum_{i=1}^{N}\mathbf{A(y}_{j},\mathbf{x}_{i})\mathbf{%
\omega }_{i}=\sum_{\mathbf{x}_{i}\in \Omega \left( \mathbf{y}_{j}\right) }%
\mathbf{A(y}_{j},\mathbf{x}_{i})\mathbf{\omega }_{i}+\sum_{\mathbf{x}%
_{i}\notin \Omega \left( \mathbf{y}_{j}\right) }\mathbf{A(y}_{j},\mathbf{x}%
_{i})\mathbf{\omega }_{i}  \label{fmm1} \\
&=&\sum_{i=1}^{N}\mathbf{A}^{(sparse)}\mathbf{(y}_{j},\mathbf{x}_{i})\mathbf{%
\omega }_{i}+\sum_{i=1}^{N}\mathbf{A}^{(dense)}\mathbf{(y}_{j},\mathbf{x}%
_{i})\mathbf{\omega }_{i},\quad j=1,...,M,  \notag
\end{eqnarray}
where $\mathbf{x}_{1},...,\mathbf{x}_{N}$ are the sources, $\mathbf{y}%
_{1},...,\mathbf{y}_{M}$ are the receivers$,$ $\Omega \left( \mathbf{y}%
_{j}\right) $ is a neighborhood of a box containing $\mathbf{y}_{j}$ which
determines the sparse and dense parts of the matrix $\mathbf{A}$. Local
summation, or sparse matrix-vector multiplication, is performed directly,
while the dense matrix-vector product is found via generation of multipole
expansions, translations, and evaluations of the local expansions. The
present paper is about an efficient way to perform the dense matrix-vector
product and does not consider acceleration of the sparse matrix-vector
multiplication.

\begin{table}[htb]
\caption{Comparison of the FMMs for the Laplace equation and for the VEM}%
\begin{tabular}{|p{1.75in}|p{1.75in}|p{2.5in}|}
\hline
FMM for Laplace equation & FMM for VEM & Comments \\ \hline
\multicolumn{3}{|c|}{\textbf{FMM set}} \\ \hline
Create data structure & The same & octree, neighbor lists, etc. \\ \hline
Precompute translations & The same & only for a single potential \\ \hline
\multicolumn{3}{|c|}{\textbf{FMM run}} \\ \hline
Create multipole expansions & Modified for two potentials & see Sec. 4.3 \\ 
\hline
M2M translations & Modified for two potentials & Sec. 4.4 \\ \hline
M2L translations & Modified for two potentials & Sec. 4.4 \\ \hline
L2L translations & Modified for two potentials & Sec. 4.4 \\ \hline
Evaluate expansions & Modified for two potentials & Sec. 4.5 \\ \hline
Local summation & Add velocity and stretching & direct evaluation of partial
sums (\ref{1}) \& (\ref{s1.1}) \\ \hline
\end{tabular}
\end{table}
\subsection{Complexity of the FMM}

The relative cost of the different steps of the FMM depends on the source
and receiver distributions, truncation number $p$, and the depth of the
octree, $l_{\max }$, which hierarchically partitions the computational
domain occupied by $N$ sources and $M$ receivers. Simple cost estimates for
the dense and sparse matrix-vector products can be provided for random
uniform distributions and $M\sim N$: 
\begin{equation}
C^{(dense)}=N\left( A_{1}+\frac{A_{2}}{s}\right) ,\quad
C^{(sparse)}=B_{1}Ns,\quad s=N\cdot 8^{-l_{\max }},  \label{fmm2}
\end{equation}%
where $A_{1}$ is the sum of costs of generation and evaluation of single
expansions, constant $A_{2}$ is determined by the cost of translations per
box, $B_{1}$ is the constant determined by the complexity of direct local
summation for a single receiver, and $s$ is the number of sources at level $%
l_{\max }$, which should be found from optimization. Note that qualitatively
different $C^{(sparse)}$ dependence on $N$ and $s$ holds for some computing
architectures, e.g. for graphics processors (GPUs) \cite{Gumerov2008:JCP}.
It is also noticeable that cost $NA_{1}$ can be neglected compared to other
costs almost in all cases (and for simplicity we neglect it as well). In any
case, theoretically, the optimum performance for a serial CPU implementation
of the FMM can be achieved when the sum of the costs as a function of $s$
reaches minimum, i.e. 
\begin{equation}
s_{opt}=\sqrt{\frac{A_{2}}{B_{1}}}\text{,\quad }C_{opt}^{(dense)}\approx
C_{opt}^{(sparse)}=N\sqrt{A_{2}B_{1}},\quad C_{opt}^{(total)}\approx 2N\sqrt{%
A_{2}B_{1}}.  \label{fmm3}
\end{equation}

Based on this we can find theoretical complexity ratios of different
versions of the FMM. As a reference we use a harmonic FMM for a single
potential, where the gradient of potential is also computed. The operation
count and our numerical experiments (see section below) show that in this
case $B_{1}$ is approximately the same for the scalar harmonic FMM and
velocity+stretching computations. Hence, only $A_{2}$ is changed in the case
of the FMM for vortex methods. Eq. (\ref{fmm3}) then shows that the cost
increase for a standard three potential representation will be $\sqrt{3}$
times, while for the two potential representation only $\sqrt{2}$ times, or
that the latter method is $\sqrt{3/2}\approx 1.22$ times faster than the
former method. Note then that in practice the depth of the octree can be
changed only discretely, and a perfect balance of the sparse and dense parts
of the algorithm cannot be achieved, so some fluctuations around the value $%
\sqrt{2}$ are expected. Analysis of efficiency of implementations with fine
and coarse grained parallelism, such as \cite{Hu2011:SC}, also can be done,
but this requires particular architecture considerations, which goes beyond
the scope of this paper.

\subsection{Modification of complex valued harmonic FMM}

It is proposed in \cite{Gumerov2006:JCP} to modify an available FMM routine
for complex valued harmonic function to an FMM routine which provides the
FMM for real valued biharmonic functions. So just one complex FMM can be
executed instead of two FMMs for the real functions. Our tests show that
such an approach provides a small advantage compared to the FMMs for real
harmonic functions. This method can be taken and applied directly to the
present case, since a complex valued harmonic function $\Psi \left( \mathbf{r%
}\right) $ can be composed from two real functions $\phi \left( \mathbf{r}%
\right) $ and $\chi \left( \mathbf{r}\right) $, as 
\begin{equation}
\Psi \left( \mathbf{r}\right) =\phi \left( \mathbf{r}\right) +i\chi \left( 
\mathbf{r}\right) .  \label{chf1}
\end{equation}
Further the translation algorithm for $\Psi \left( \mathbf{r}\right) $ will
be exactly the same as for the biharmonic functions, described in \cite%
{Gumerov2006:JCP}, with the only difference in the conversion operators.

\subsection{Generation of multipole expansions}

Here we propose two methods to generate multipole expansions. The first
method requires just a slight modification of a function generating the
multipole expansion of a dipole source, which are available in many harmonic
FMM codes. The second method utilizes a function generating the multipole
expansion of a monopole source.

\subsubsection{Method 1}

Consider multipole expansion of $\mathbf{v}_{l}\left( \mathbf{y}\right) $
given by Eq. (\ref{1}) about the center, $\mathbf{x}_{\ast },$of a source
box $b$ containing $\mathbf{x}_{l}$ and denote\ $\mathbf{r}=\mathbf{y-x}%
_{\ast },$ $\mathbf{r}_{l}=\mathbf{x}_{l}-\mathbf{x}_{\ast }$. The purpose
is to find coefficients of scalar potentials $\phi _{ln}^{m}$ and $\chi
_{ln}^{m}$, which then should be summed up with respect to $l$, $\mathbf{x}%
_{l}\in b$ to obtain coefficients $\phi _{(b)n}^{m}$ and $\chi _{(b)n}^{m}$
for the box, which further should be used in the translation process.

It is not difficult to show that the auxiliary harmonic functions 
\begin{equation}
\psi _{l}=\mathbf{r}\cdot \nabla \phi _{l}=\mathcal{D}_{\mathbf{r}}\phi
_{l},\quad \varpi _{l}=\chi _{l}+\mathbf{r}\cdot \nabla \chi _{l}=\left( 
\mathcal{I}+\mathcal{D}_{\mathbf{r}}\right) \chi _{l},  \label{eve5}
\end{equation}%
are dipoles, 
\begin{equation}
\psi _{l}=\mathbf{r}\cdot \nabla \times \frac{\boldsymbol{\omega }_{l}}{%
\left\vert \mathbf{r}-\mathbf{r}_{l}\right\vert }=\left( \mathbf{r}%
_{l}\times \boldsymbol{\omega }_{l}\right) \mathbf{\cdot }\left( \frac{%
\mathbf{r-r}_{l}}{\left\vert \mathbf{r}-\mathbf{r}_{l}\right\vert ^{3}}%
\right) ,  \label{eve7}
\end{equation}%
\begin{equation}
\varpi _{l}=\nabla \cdot \frac{\mathbf{\omega }_{l}}{\left\vert \mathbf{r}-%
\mathbf{r}_{l}\right\vert }=\frac{-\boldsymbol{\omega }_{l}\cdot \left( 
\mathbf{r}-\mathbf{r}_{l}\right) }{\left\vert \mathbf{r}-\mathbf{r}%
_{l}\right\vert ^{3}},  \label{eve9}
\end{equation}%
i.e. $\psi _{l}$ is a dipole with moment $\mathbf{p}_{l}=\mathbf{r}%
_{l}\times \boldsymbol{\omega }_{l}$, while $\varpi _{l}$ is a dipole with
moment $\mathbf{q}_{l}=$ $-\mathbf{\omega }_{l}.$ Hence, coefficients $\psi
_{ln}^{m}$ and $\varpi _{ln}^{m}$ can be found using the dipole expansion
procedure. We can determine coefficients $\phi _{ln}^{m}$
and $\chi _{ln}^{m}$, as the operators in Eq. (\ref{eve5}) are diagonal in
the $S$ basis (see Eqs (\ref{dr2}) and (\ref{dr4})): 
\begin{equation}
\phi _{ln}^{m}=-\frac{1}{n+1}\psi _{ln}^{m},\quad \chi _{ln}^{m}=-\frac{1}{n}%
\varpi _{ln}^{m},\quad \left( \chi _{l0}^{0}=0\right) .  \label{eve10}
\end{equation}

\subsubsection{Method 2}

Using Eq. (\ref{rsbf5}), we obtain 
\begin{equation}
\mathbf{v}_{l}\left( \mathbf{y}\right) =\nabla \times \frac{%
\boldsymbol{\omega }_{l}}{\left\vert \mathbf{r}-\mathbf{r}_{l}\right\vert }%
=\sum_{n=0}^{\infty }\sum_{m=-n}^{n}g_{ln}^{m}\mathbf{F}_{ln}^{m}\left( 
\mathbf{r}\right) ,\quad \mathbf{F}_{ln}^{m}\left( \mathbf{r}\right) =\nabla
\times \left[ \mathbf{\omega }_{l}S_{n}^{m}(\mathbf{r})\right] ,\quad
g_{ln}^{m}=R_{n}^{-m}(-\mathbf{r}_{l}).  \label{eve1}
\end{equation}%
Note then that $\mathbf{F}_{ln}^{m}$ is equivalent to the right hand side of
Eq. (\ref{CVL6}), where one should set $\mathbf{t=\omega }_{l}$, $\widehat{%
\chi }=S_{n}^{m}(\mathbf{r}).$ So this function can be represented in the
form provided by the left hand side of Eq. (\ref{CVL6}), where functions $%
\phi ^{\prime }$ and $\chi ^{\prime }$ can be found from Eqs (\ref{CVL11})
and (\ref{CVL14}) i.e. 
\begin{eqnarray}
\mathbf{F}_{ln}^{m}\left( \mathbf{r}\right)  &=&\nabla \times \left[ \mathbf{%
\omega }_{l}S_{n}^{m}(\mathbf{r})\right] =\nabla \Phi _{ln}^{m}\left( 
\mathbf{r}\right) +\nabla \times \left( \mathbf{r}X_{ln}^{m}\left( \mathbf{r}%
\right) \right) ,  \label{eve2} \\
\Phi _{ln}^{m}\left( \mathbf{r}\right)  &=&\mathcal{C}_{12}\left( \mathbf{%
\omega }_{l}\right) S_{n}^{m}(\mathbf{r}),\quad X_{ln}^{m}\left( \mathbf{r}%
\right) =\left( \mathcal{C}_{22}\left( \boldsymbol{\omega }_{l}\right) -%
\mathcal{I}\right) S_{n}^{m}(\mathbf{r}).  \notag
\end{eqnarray}%
Substituting this into Eq. (\ref{eve1}) and using representation of the
conversion operators in the $S$ basis, Eq. (\ref{drt11}), we obtain 
\begin{eqnarray}
\mathbf{v}_{l}\left( \mathbf{y}\right)  &=&\nabla \phi _{l}\left( \mathbf{r}%
\right) +\nabla \times \left( \mathbf{r}\chi _{l}\left( \mathbf{r}\right)
\right) ,  \label{eve3} \\
\phi _{l}\left( \mathbf{r}\right)  &=&\sum_{n=0}^{\infty
}\sum_{m=-n}^{n}g_{ln}^{m}\mathcal{C}_{12}\left( \boldsymbol{\omega }%
_{l}\right) S_{n}^{m}(\mathbf{r})=\sum_{n=0}^{\infty }\sum_{m=-n}^{n}\phi
_{ln}^{m}S_{n}^{m}(\mathbf{r}),\quad   \notag \\
\chi _{l}\left( \mathbf{r}\right)  &=&\sum_{n=0}^{\infty
}\sum_{m=-n}^{n}g_{ln}^{m}\left( \mathcal{C}_{22}\left( \boldsymbol{\omega }%
_{l}\right) -\mathcal{I}\right) S_{n}^{m}(\mathbf{r})=\sum_{n=0}^{\infty
}\sum_{m=-n}^{n}\chi _{ln}^{m}S_{n}^{m}(\mathbf{r}),\quad   \notag
\end{eqnarray}%
where 
\begin{eqnarray}
\phi _{ln}^{m} &=&\frac{1}{n+1}\left[ (\omega _{lx}-i\omega _{ly})\frac{n-m+1%
}{2}g_{ln}^{m-1}-(\omega _{lx}+i\omega _{ly})\frac{n+m+1}{2}%
g_{ln}^{m+1}-i\omega _{lz}mg_{ln}^{m}\right] ,  \label{eve4} \\
\chi _{ln}^{m} &=&-\frac{1}{n}\left[ \frac{1}{2}(\omega _{ly}+i\omega
_{lx})g_{l,n-1}^{m-1}-\frac{1}{2}(\omega _{ly}-i\omega
_{lx})g_{l,n-1}^{m+1}-\omega _{lz}g_{l,n-1}^{m}\right] ,\quad \left( \chi
_{l0}^{0}=0\right) .  \notag
\end{eqnarray}

\subsection{Translations}

\begin{figure}[tbh]
\begin{center}
\includegraphics[width=5.5in,trim=0 7in 0 0.8in]
{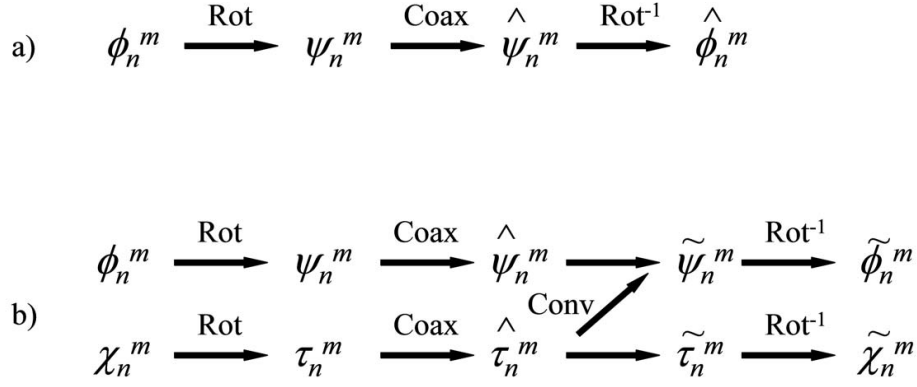}
\end{center}
\caption{Translation schemes for the FMM for the scalar Laplace equation (a),
and the FMM for the DCVLE (b), based on the RCR-decomposition. The operators are shown abbreviated as follows: Rot: rotation, Coax: coaxial translation, Rot$^{-1}$: back (inverse) rotation. }
\label{FigTscheme}
\end{figure}

In many FMM codes translations are performed using the rotation-coaxial
translation-back rotation (RCR) decomposition of the translation operators,
which reduces translation cost of expansions of length $p^{2}$ to $O(p^{3}),$
opposed to $O(p^{4})$ required for the direct application of the translation
matrix (e.g. see \cite{Gumerov2006:JCP}). Such a decomposition is also
beneficial for faster conversion, since the rotations do not change the form
of decomposition of the vector field (\ref{4}) and there is no need to
rotate $\phi $ and $\chi $ in conversion operators. Coaxial translation
means translation along the $z$ direction to distance $t$, in which case
expressions for the conversion operators (\ref{drt10}) and (\ref{drt11})
become even simpler ($\mathbf{t}=t\mathbf{i}_{z},$ $t_{x}=t_{y}=0,$ $t_{z}=t$%
): 
\begin{eqnarray}
R\text{ conversion} &:&\text{\quad }\widetilde{\phi }_{n}^{m}=\widehat{\phi }%
_{n}^{m}+it\frac{m}{n}\widehat{\chi }_{n}^{m},\quad \left( \widetilde{\phi }%
_{0}^{0}=\widehat{\phi }_{0}^{0}\right) ,\quad \widetilde{\chi }_{n}^{m}=%
\widehat{\chi }_{n}^{m}-\frac{t}{n+1}\widehat{\chi }_{n+1}^{m},  \label{rcr1}
\\
S\text{ conversion} &:&\text{\quad }\widetilde{\phi }_{n}^{m}=\widehat{\phi }%
_{n}^{m}-it\frac{m}{n+1}\widehat{\chi }_{n}^{m},\quad \widetilde{\chi }%
_{n}^{m}=\widehat{\chi }_{n}^{m}+\frac{t}{n}\widehat{\chi }_{n-1}^{m},\quad
\left( \widetilde{\chi }_{0}^{0}=\widehat{\chi }_{0}^{0}\right) .  \notag
\end{eqnarray}%
Figure \ref{FigTscheme} illustrates the present translation scheme which
uses the RCR-decomposition (rotation-coaxial translation-back rotation). 

\subsection{Evaluation of local expansions}

\subsubsection{Velocity}

As a result of the FMM downward pass the $R$ expansions of scalar potentials
are obtained about the center $\mathbf{y}_{\ast },$of an evaluation box $b$
containing receiver point $\mathbf{y}$%
\begin{equation}
\phi \left( \mathbf{r}\right) =\sum_{n=0}^{\infty }\sum_{m=-n}^{n}\phi
_{n}^{m}R_{n}^{m}(\mathbf{r}),\quad \chi \left( \mathbf{r}\right)
=\sum_{n=0}^{\infty }\sum_{m=-n}^{n}\chi _{n}^{m}R_{n}^{m}(\mathbf{r}),\quad 
\mathbf{r}=\mathbf{y-y}_{\ast }.  \label{evf1}
\end{equation}%
Cartesian components of the velocity can be obtained by projection of Eq. (%
\ref{4}) to the basis vectors $\mathbf{i}_{x},\mathbf{i}_{y},$ and $\mathbf{i%
}_{z}$ as follows 
\begin{equation}
v_{k}=\mathbf{i}_{k}\cdot \mathbf{v=i}_{k}\cdot \nabla \phi +\mathbf{i}%
_{k}\cdot \nabla \times \left( \mathbf{r}\chi \right) =\mathcal{D}_{\mathbf{i%
}_{k}}\phi +\mathcal{D}_{\mathbf{r\times i}_{k}}\chi ,\quad k=x,y,z.
\label{evf2}
\end{equation}%
since $\mathbf{i}_{k}\cdot \nabla \times \left( \mathbf{r}\chi \right) =%
\mathbf{i}_{k}\cdot \left( \nabla \chi \times \mathbf{r}\right) =\left( 
\mathbf{r\times i}_{k}\right) \cdot \nabla \chi $. Using representations of
the above operators in the $R$ basis, (\ref{dt3}) and (\ref{drt7}), where $%
\mathbf{t}=\mathbf{i}_{k}$, we determine 
\begin{eqnarray}
v_{k} &=&\sum_{n=0}^{\infty }\sum_{m=-n}^{n}v_{kn}^{m}R_{n}^{m}(\mathbf{r}%
),\quad k=x,y,z,  \label{evf3} \\
v_{xn}^{m} &=&\frac{1}{2}\left[ i\phi _{n+1}^{m-1}+i\phi _{n+1}^{m+1}+\left(
n-m\right) \chi _{n}^{m+1}-\left( n+m\right) \chi _{n}^{m-1}\right] ,  \notag
\\
v_{yn}^{m} &=&\frac{1}{2}\left[ \phi _{n+1}^{m-1}-\phi _{n+1}^{m+1}+i\left(
n-m\right) \chi _{n}^{m+1}+i\left( n+m\right) \chi _{n}^{m-1}\right] , 
\notag \\
v_{zn}^{m} &=&-\phi _{n+1}^{m}-im\chi _{n}^{m}.  \notag
\end{eqnarray}

\subsubsection{Stretching term and strain tensor}

Furthermore, consider computation of the vortex stretching at evaluation
point $\mathbf{y}_{j}$ $\left( \mathbf{r}_{j}=\mathbf{y}_{j}\mathbf{-y}%
_{\ast }\right) $, assuming that the strength vector at this point is $%
\boldsymbol{\omega }_{j}$. The stretching is a vector 
\begin{equation}
\mathbf{s}_{j}=\left. \left( \boldsymbol{\omega }_{j}\cdot \nabla \right) 
\mathbf{v}\left( \mathbf{r}\right) \right| _{\mathbf{r=r}_{j}}=\left. 
\mathcal{D}_{\boldsymbol{\omega }_{j}}\mathbf{v}\left( \mathbf{r}\right)
\right| _{\mathbf{r=r}_{j}}=\sum_{k}\mathbf{i}_{k}\left. \mathcal{D}_{%
\boldsymbol{\omega }_{j}}v_{k}\left( \mathbf{r}\right) \right| _{\mathbf{r=r}%
_{j}}.  \label{evf4}
\end{equation}
Hence, the Cartesian components of this vector can be obtained simply from
computed coefficients $v_{kn}^{m}$, Eq. (\ref{evf3}), to which sparse
operator $\mathbf{D}_{\boldsymbol{\omega }_{j}}^{(R)}$ should be applied
(see Eq. (\ref{dt3})): 
\begin{eqnarray}
s_{jk} &=&\sum_{n=0}^{\infty }\sum_{m=-n}^{n}s_{jkn}^{m}R_{n}^{m}(\mathbf{r}%
_{j}),\quad k=x,y,z,  \label{evf5} \\
s_{jkn}^{m} &=&\frac{1}{2}\left[ \left( \omega _{jy}+i\omega _{jx}\right)
v_{k,n+1}^{m-1}-(\omega _{jy}-i\omega _{jx})v_{k,n+1}^{m+1}\right] -\omega
_{jz}v_{k,n+1}^{m}.  \notag
\end{eqnarray}
In practice, it is more efficient to compute expansion coefficients $%
u_{lkn}^{m}$ for the functions $u_{lk}=\mathcal{D}_{\mathbf{i}_{l}}v_{k},$ $%
l,k=1,2,3,$ which do not depend on the evaluation point $j$ and then form the coefficients $s_{jkn}^{m}$ for each point as 
\begin{equation}
s_{jkn}^{m}=\sum_{l=1}^{3}\omega _{jl}u_{lkn}^{m}.  \label{evf6}
\end{equation}
Note that $u_{lk}$ are components of the tensor $\nabla \mathbf{v}$. The contraction $\beta =\nabla \mathbf{v:}\nabla 
\mathbf{v}$ can then be computed (for compressible flows). Also,
computation of $\nabla \mathbf{v}$ provides the strain tensor, which can be
used for modeling complex fluids.

\section{Numerical tests}

For numerical tests we used our FMM software which employs
RCR-decomposition of translation operators, and modified it for two harmonic
functions. Conversion operators in the $R$ or $S$ basis were executed after
coaxial translation operators, as shown in Fig. \ref{FigTscheme}. Additional
small modifications were used in the algorithm to compute $R$ basis
functions for real harmonic functions recursively, as presented in \cite%
{Gumerov2008:JCP}. In contrast to \cite{Gumerov2008:JCP} no optimizations of
the algorithm were used (no GPU acceleration, standard 189 M2L translation stencils, no
variable truncation number, etc.), as the main purpose of this paper was to
provide a basic comparative performance and accuracy test of the method, with a basic implementation.
Open MP parallelization was used, which for a 4 core PC provided
parallelization efficiency close to 100 percent. Wall clock times reported
below were measured on an Intel QX6780 (2.8 GHz) 4 core PC with 8GB RAM.

\subsection{Error tests}

The first test we conducted is related to the numerical errors of
computation of the velocity and stretching term. Also for comparisons we
executed the FMM for a single harmonic function and measured numerical
errors in the evaluation of the potential and its gradient. There are two
basic sources of errors. The first one is due to truncation of the infinite
series. These errors are controlled by the truncation number $p$ (the
infinite series (\ref{rsbf6}) were replaced by the first $p^{2}$ terms, $%
n=0,...,p-1;$ $m=-n,...,n$), which we varied in the tests. The second source
of errors is due to the roundoff, which in our computations with double
precision in the range of tested $p$ were smaller than the truncation errors
(the roundoff errors were observed for $p\gtrsim 25$). The basic test was
performed for $N$ sources/receivers distributed randomly and uniformly
inside a cube. The error, $\epsilon _{2}$, was measured in the $L_{2}$
relative norm based on 1000 points randomly selected from the source set.
(Our previous tests using direct computations show that even 100 points
provide sufficient confidence for the $L_{2}$-norm error, see \cite%
{Gumerov2006:JCP}.) For the reference solution the velocity field,
stretching term, potential and gradient were computed directly.

\begin{figure}[htb]
\begin{center}
\includegraphics[scale=1.5,width=5.5in, trim=0 0 0 0.6in]
{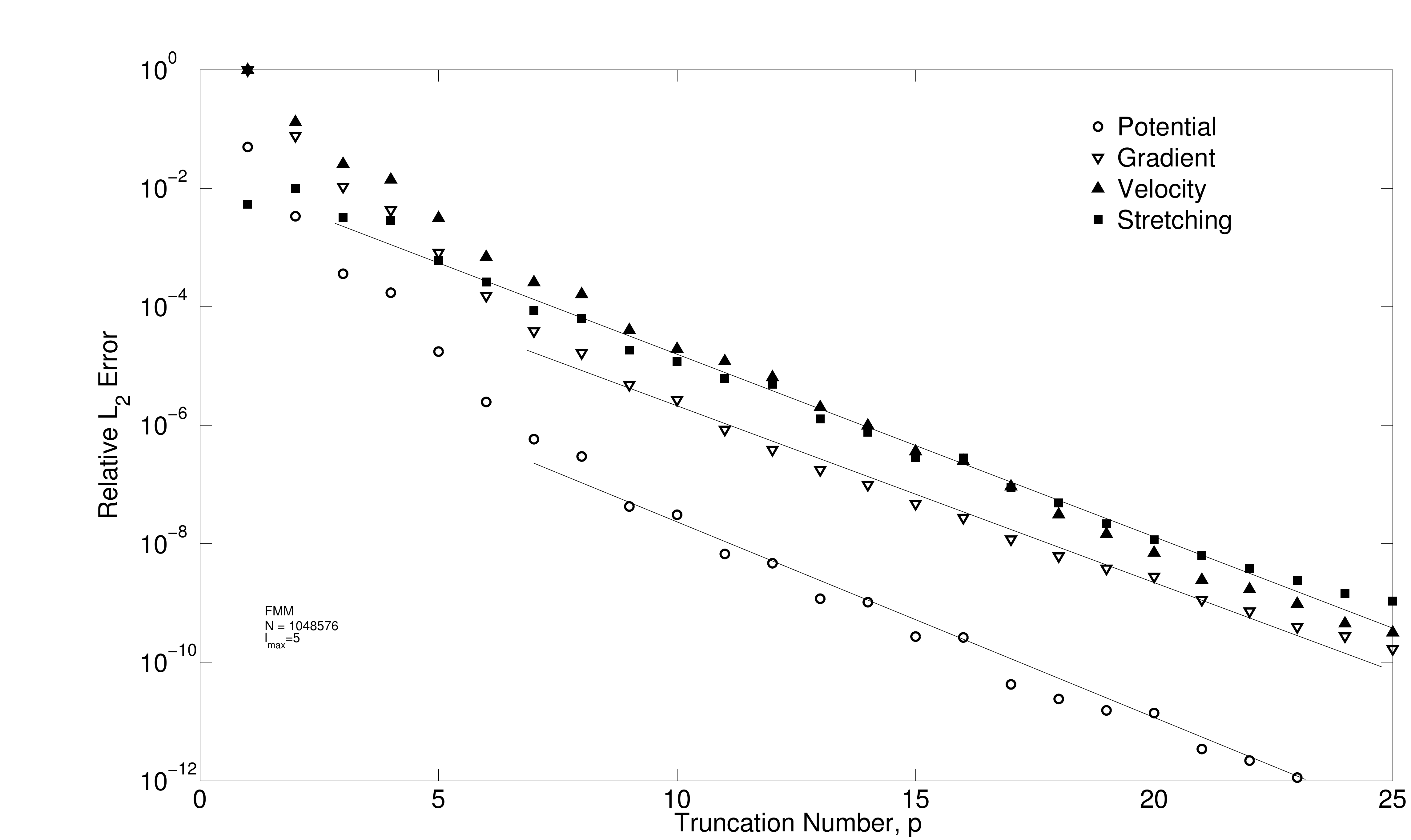}
\end{center}
\caption{Dependences of the relative FMM errors in the $L_{2}$ norm on the
truncation number, $p$. Errors were computed over 1000 random points for $%
N=2^{20}$ sources of random intensity distributed uniformly randomly inside
a cube. The maximum level of space subdivision $l_{\max }=5$. }
\label{Fig1}
\end{figure}

Figure \ref{Fig1} illustrates behavior of the computed errors for the
velocity and stretching term. For reference, the dependence of the
respective errors in the harmonic potential and in its gradient are also
shown. It is seen that starting with $p\approx 7$ spectral convergence is
observed for all cases. It is also noticeable, that the errors in potential
computations are substantially smaller than that for the gradient or higher
derivative computations. There are two basic reasons for this. First, the
effective truncation number for each derivative is smaller by one compared
to the potential, and second that in the truncated term for the derivative
an additional factor $\sim p$ appears.

\subsection{Performance tests}

For optimal FMM performance the depth of the octree $l_{\max }$ should be
chosen to minimize the total execution time. For all reported test cases we
conducted such an optimization. Some results of the profiling (wall clock
time in seconds) with random uniform distributions of sources inside a cube
and on the surface of a sphere are provided in Tables 1 and 2. In these
tables $\mathbf{v}$ and $\mathbf{s}$ indicate computations of the velocity
and stretching term for vortical flows, while $\phi $ and $\nabla \phi $
refer to a reference case for the scalar Laplace equation, where the
potential or both potential and its gradient should be computed. The total
initialization time, which includes the data structure and precomputations
related to translation operators can be amortized for a constant
source/receiver set, and is reported separately from the total run time. As
one can see this time is relatively small, while for dynamic problems it
should be added to the total run time.

The times for the local sum (``sparse'') and far field (``dense'')
matrix-vector products in the FMM are also reported. The latter is also
expanded to show timing of the FMM stages. The truncation number for all
cases was $p=12$, which provides errors $\epsilon _{2}\sim 10^{-5}$ for the
velocity and stretching term computations, while smaller errors for $\phi $
and $\nabla \phi $ (see Fig. \ref{Fig1}).

\begin{table}[htb]
\caption{ Profiling of the FMM for random uniform distribution of sources
inside a cube, $p=12$.}
\label{Table1}%
\begin{centering}
\begin{small}
\begin{tabular}{|l|l|l|l|l|l|l|l|l|}
\hline \hline
Case & $l_{\max }$ & Total Init & S-expansion & Upward & Downward & 
R-evaluation & Sparse MV & Total Run \\ \hline 
\multicolumn{9}{|c|}{$N=2^{19}$}  \\  \hline
$\mathbf{v}$ and $\mathbf{s}$ & 4 & 0.55 & 0.55 & 0.04 & 2.65 & 0.52 & 25.9
& 29.7 \\ \hline
$\mathbf{v}$ alone & 4 & 0.55 & 0.55 & 0.04 & 2.65 & 0.34 & 16.4 & 20.0 \\ \hline
$\phi $ and $\nabla \phi $ & 5 & 1.20 & 0.20 & 0.12 & 10.7 & 0.36 & 3.95 & 
15.3 \\ \hline
$\phi $ alone & 5 & 1.20 & 0.20 & 0.12 & 10.7 & 0.21 & 1.89 & 13.1 \\ \hline
\multicolumn{9}{|c|}{$N=2^{20}$}   \\ \hline
$\mathbf{v}$ and $\mathbf{s}$ & 5 & 1.71 & 1.05 & 0.27 & 25.3 & 1.11 & 14.2
& 41.9 \\ \hline
$\mathbf{v}$ alone & 5 & 1.71 & 1.05 & 0.27 & 25.3 & 0.59 & 9.04 & 36.3 \\ \hline
$\phi $ and $\nabla \phi $ & 5 & 1.71 & 0.39 & 0.12 & 10.7 & 0.71 & 15.4 & 
27.3 \\ \hline
$\phi $ alone & 5 & 1.71 & 0.39 & 0.12 & 10.7 & 0.43 & 7.32 & 19.0\\ \hline
\hline
\end{tabular}
\end{small}
\end{centering}
\end{table}

\begin{table}[tbh]
\caption{Profiling of the FMM for random uniform distribution of sources on
a sphere surface, $p=12$. }
\label{Table2}%
\begin{centering}
\begin{small}
\begin{tabular}{|l|l|l|l|l|l|l|l|l|}\hline \hline
Case & $l_{\max }$ & Total Init & S-expansion & Upward & Downward & 
R-evaluation & Sparse MV & Total Run \\ \hline
\multicolumn{9}{|c|}{$N=2^{19}$}  \\ \hline
$\mathbf{v}$ and $\mathbf{s}$ & 7 & 1.23 & 0.88 & 0.34 & 9.48 & 0.66 & 3.75
& 15.1 \\ \hline
$\mathbf{v}$ alone & 6 & 0.66 & 0.63 & 0.10 & 2.46 & 0.34 & 8.66 & 12.2 \\ \hline
$\phi $ and $\nabla \phi $ & 7 & 1.23 & 0.23 & 0.15 & 3.93 & 0.41 & 3.90 & 
8.62 \\ \hline
$\phi $ alone & 7 & 1.23 & 0.23 & 0.15 & 3.93 & 0.26 & 1.86 & 6.43 \\ \hline
\multicolumn{9}{|c|}{$N=2^{20}$} \\ \hline
$\mathbf{v}$ and $\mathbf{s}$ & 7 & 1.81 & 1.29 & 0.34 & 9.48 & 1.30 & 14.1
& 26.5 \\ \hline
$\mathbf{v}$ alone & 7 & 1.81 & 1.29 & 0.34 & 9.48 & 0.70 & 8.86 & 20.7 \\ \hline
$\phi $ and $\nabla \phi $ & 7 & 1.81 & 0.46 & 0.15 & 4.14 & 0.84 & 15.3 & 
20.9 \\ \hline
$\phi $ alone & 7 & 1.81 & 0.46 & 0.15 & 4.14 & 0.52 & 7.30 & 12.6\\ \hline
\hline
\end{tabular}
\end{small}
\end{centering}
\end{table}

\begin{figure}[tb]
\begin{center}
\includegraphics[width=5.5in, trim=1in 0 0 0]
{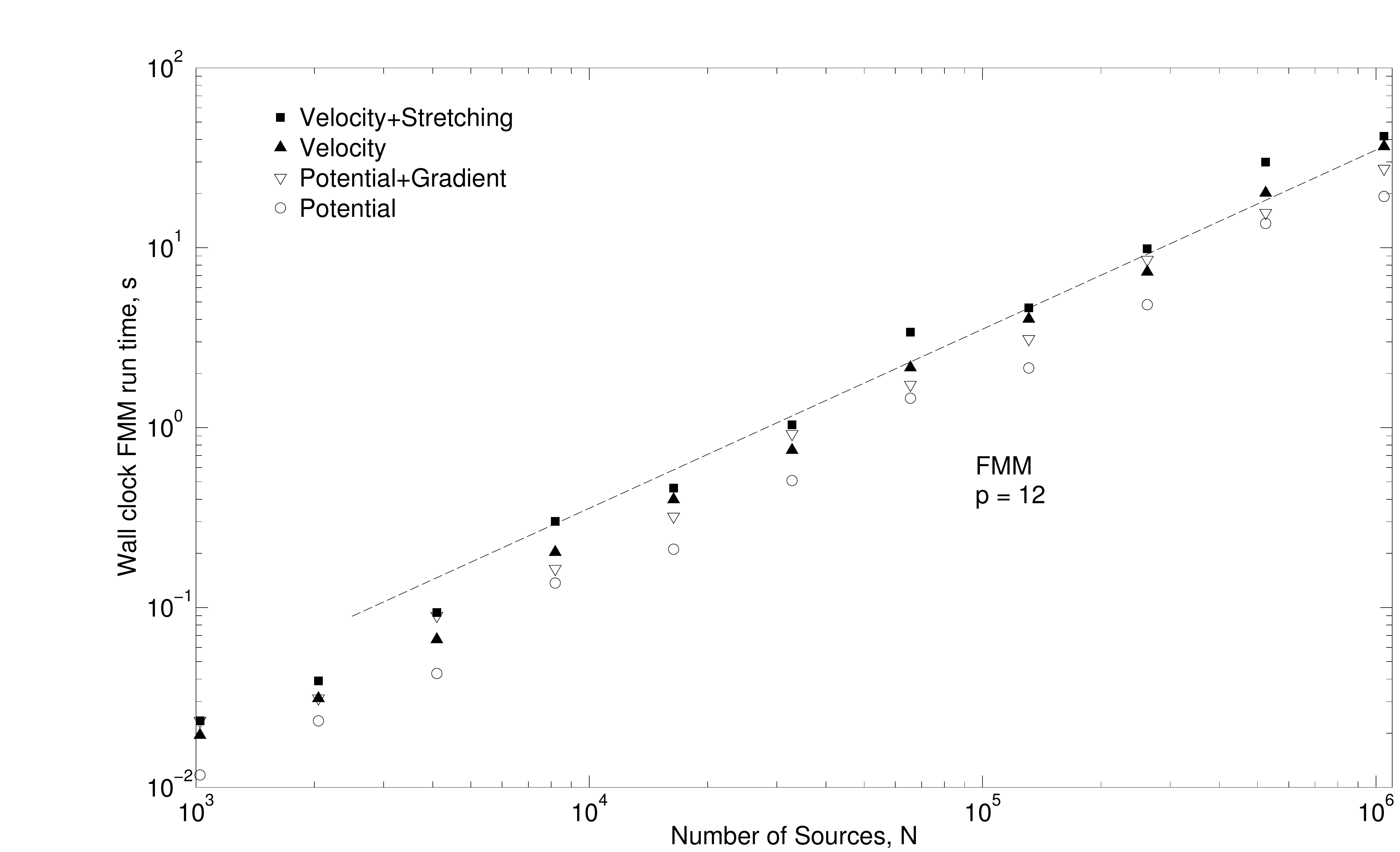}
\end{center}
\caption{The wall clock time for the FMM for computation of vortical and
potential flows (different terms and compinations). The straight line shows
linear dependence. For all cases the sources are distributed randomly and
uniformly inside a cube; the truncation number is constant, $p=12$.}
\label{Fig2}
\end{figure}
\begin{figure}[tbh]
\begin{center}
\includegraphics[width=5.5in,trim=1in 0 0 1in]
{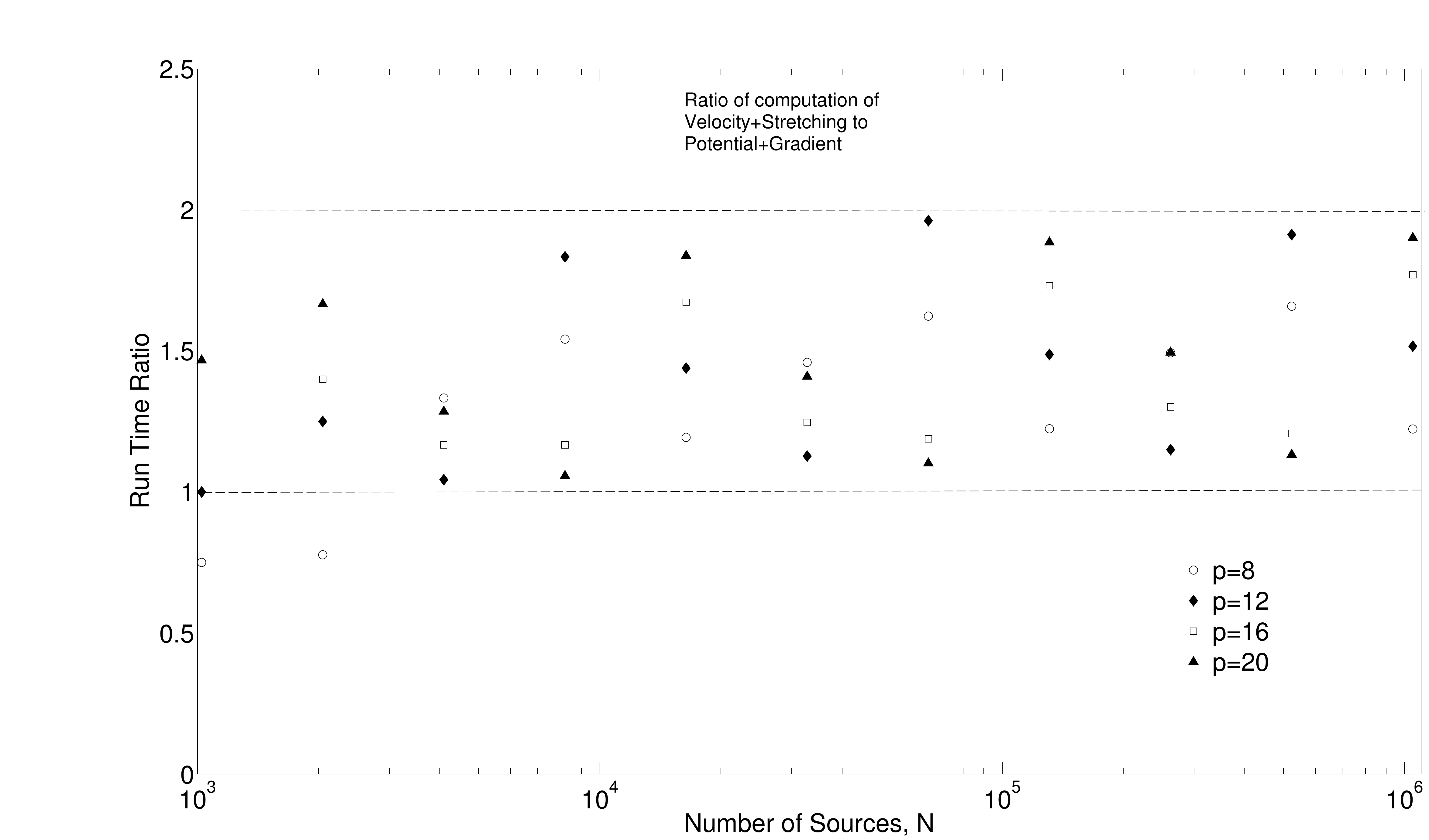}
\end{center}
\caption{The ratio of the FMM run time for computation of the velocity and
stretching term in a vortical flow to the respective time for the potential
and gradient computations in a potential flow for different truncation
numbers $p$ and number of sources, $N$, which were randomly and uniformly
distributed inside a cube. All data for $N>2048$ are located between the
dashed lines.}
\label{Fig3}
\end{figure}

The tables show that in the cases when the number of translations for a
single potential $\phi $ for the scalar Laplace equation and coupled
potentials $\phi $ and $\chi $ for the DCVLE are the same (the same $l_{\max
})$ the translation time for the latter case is approximately twice, as
expected. Deviations may be explained by two factors. First, this is due to
increase in the size of the arrays representing expansions and more time
needed for data access, and, second, by the presence of the conversion
operators. The tables also show that the time for sparse matrix-vector
products for velocity only computations in DCVLE is slightly larger than for
potential only computations in a harmonic FMM, while the time for the same
operations for velocity and stretching computations are slightly smaller
than for potential and gradient computations. Note, however, that if an
additional near-field kernel should be computed, which may involve
computation of special functions (exponents, error integrals, etc.) the time
for the sparse matrix-vector product would increase, while the translation
part would not be affected. Also note that, theoretically, in the optimized
algorithm, an increase of the complexity of the sparse matrix-vector product
by a factor of $k$ affects the total complexity as $\sqrt{k}$ (see Eq. (\ref%
{fmm3})). The ratio of the total time for the velocity and stretching
computations to the time of potential and gradient computations depends on
the problem. In all our numerical experiments this ratio never exceeded 2
(except for one outlier at $N=1024$, see Fig. \ref{Fig3}). 

Figure \ref{Fig2} illustrates dependence of the wall clock time on the
number of sources $N$, which in all cases was set to be equal to the number
of receivers. It is seen that at large $N$ the algorithm scales linearly,
and the time for velocity and stretching term computations is always larger
than that for scalar potential only computations by a factor of
approximately two. Figure \ref{Fig3} illustrates the wall clock FMM run time
ratio of velocity and stretching to potential and gradient computations for
different $p$ and $N=2^{k}$, $k=10,...,20$. It is seen that for $k>11$ this
ratio is larger than unity and smaller than two with a mean value about
1.405, while the average over all points shown in this figure is 1.423,
which agrees well with theoretical value $\sqrt{2}$ (see discussion after
Eq. (\ref{fmm3})). Finally, we can see that the velocity and stretching
computations are approximately twice as expensive as computation for a
single harmonic potential. This also agrees well with the theory, since the
cost of sparse matrix-vector product for a single harmonic potential
approximately two times smaller than that for the velocity and stretching
computations. Hence, in Eq. (\ref{fmm3}) both coefficients $A_{2}$ and $%
B_{1} $ for the latter case are two times larger than that for the former
case, which provides exactly factor 2.

\subsection{Example vortex computations}

We implemented vortex particle and vortex filament methods (VPM and VFM,
respectively) accelerated by the FMM described above. Some results for test
problems related to the vortex ring dynamics are presented below.

In the VFM the total velocity field is a superposition of the Biot-Savart
contour integrals taken along the vortex filaments. The field of a single
filament, $C$, can be discretized as 
\begin{eqnarray}
\mathbf{v}\left( \mathbf{y}\right) &=&\frac{1}{4\pi }\int_{C}\frac{\Gamma
\left( \mathbf{x}\right) d\mathbf{l}\left( \mathbf{x}\right) \times \left( 
\mathbf{y-x}\right) }{\left| \mathbf{y-x}\right| ^{3}}=\sum_{i=1}^{N}\mathbf{%
v}_{i}\left( \mathbf{y}\right) ,  \label{dis1} \\
\mathbf{v}_{i}\left( \mathbf{y}\right) &=&\frac{\Gamma _{i}}{4\pi }%
\int_{C_{i}}\frac{d\mathbf{l}\left( \mathbf{x}\right) \times \left( \mathbf{%
y-x}\right) }{\left| \mathbf{y-x}\right| ^{3}},\quad C=\cup _{i=1}^{N}C_{i},
\notag
\end{eqnarray}
where $N$ is the number of elements and each element $C_{i}$ can be assumed
a line segment of constant circulation $\Gamma _{i}$, in which case the
integrals can be computed analytically (see Appendix B), 
\begin{equation}
\mathbf{v}_{i}\left( \mathbf{y}\right) =\frac{\Gamma _{i}}{4\pi }\frac{%
\mathbf{r}_{i}^{(1)}\times \mathbf{r}_{i}^{(2)}}{r_{i}^{(1)}r_{i}^{(2)}+%
\mathbf{r}_{i}^{(1)}\cdot \mathbf{r}_{i}^{(2)}}\left( \frac{1}{r_{i}^{(1)}}+%
\frac{1}{r_{i}^{(2)}}\right) ,\quad \mathbf{r}_{i}^{(j)}=\mathbf{y-x}%
_{i}^{(j)},\quad r_{i}^{(j)}=\left| \mathbf{r}_{i}^{(j)}\right|, \quad \text{ 
}j=1,2,  \label{dis2}
\end{equation}
where $\mathbf{x}_{i}^{(j)}$, $j=1,2$, are coordinates of the end points of $%
C_{i}$ and condition $\mathbf{v}_{i}\left( \mathbf{y}\right) $ $=\mathbf{0}$
is imposed for $\mathbf{y}\in C_{i}$. Multipole expansion in terms of
potentials $\phi $ and $\chi $ can be found using quadrature (see Appendix
B). In the VFM vortex stretching occurs naturally since the ends of each
segment propagate with the velocity of fluid particles at those points (e.g.
see \cite{Cottet2000:Book}). Hence in the FMM for the VFM the element
centers are considered as ``sources'', while the end points are used as the
``receivers''.

For numerical examples we used the following smoothing kernel 
\begin{equation}
K(r;a)=\func{erf}\frac{r}{a\sqrt{2}}-\sqrt{\frac{2}{\pi }}\frac{r}{a}\exp
\left( -\frac{r^{2}}{2a^{2}}\right) ,  \label{vem1}
\end{equation}
which corresponds to the Gaussian vortex blob function with standard
deviation $a$.

\subsubsection{Single vortex ring computations}

The self-induced motion of a vortex ring of radius $R$ in an inviscid
incompressible fluid is a classical solution \cite{Helmholtz1867:PhilMag,
Lamb1932}. A characteristic of this problem is that the velocity of the ring
itself should be infinitely large in this case, and the way to fix this is
to introduce a small vortex core of size $\delta $, in which case 
\begin{equation}
V^{(self)}\sim \frac{\Gamma }{4\pi R}\left( \ln \frac{8R}{\delta }-\frac{1}{4%
}\right) ,\quad \delta \rightarrow 0.  \label{dis4}
\end{equation}
The asymptotic behavior of the VFM scheme at large $N$ can be estimated
theoretically (see Appendix B). 
\begin{equation}
V_{\epsilon }^{(self)}\sim \frac{\Gamma }{4\pi R}\ln \frac{2N}{\pi },\quad
N\rightarrow \infty .  \label{dis6}
\end{equation}

\begin{figure}[tbh]
\begin{center}
\includegraphics[width=5.5in, scale =1.5, trim = 4in 0 0 0.3in]
{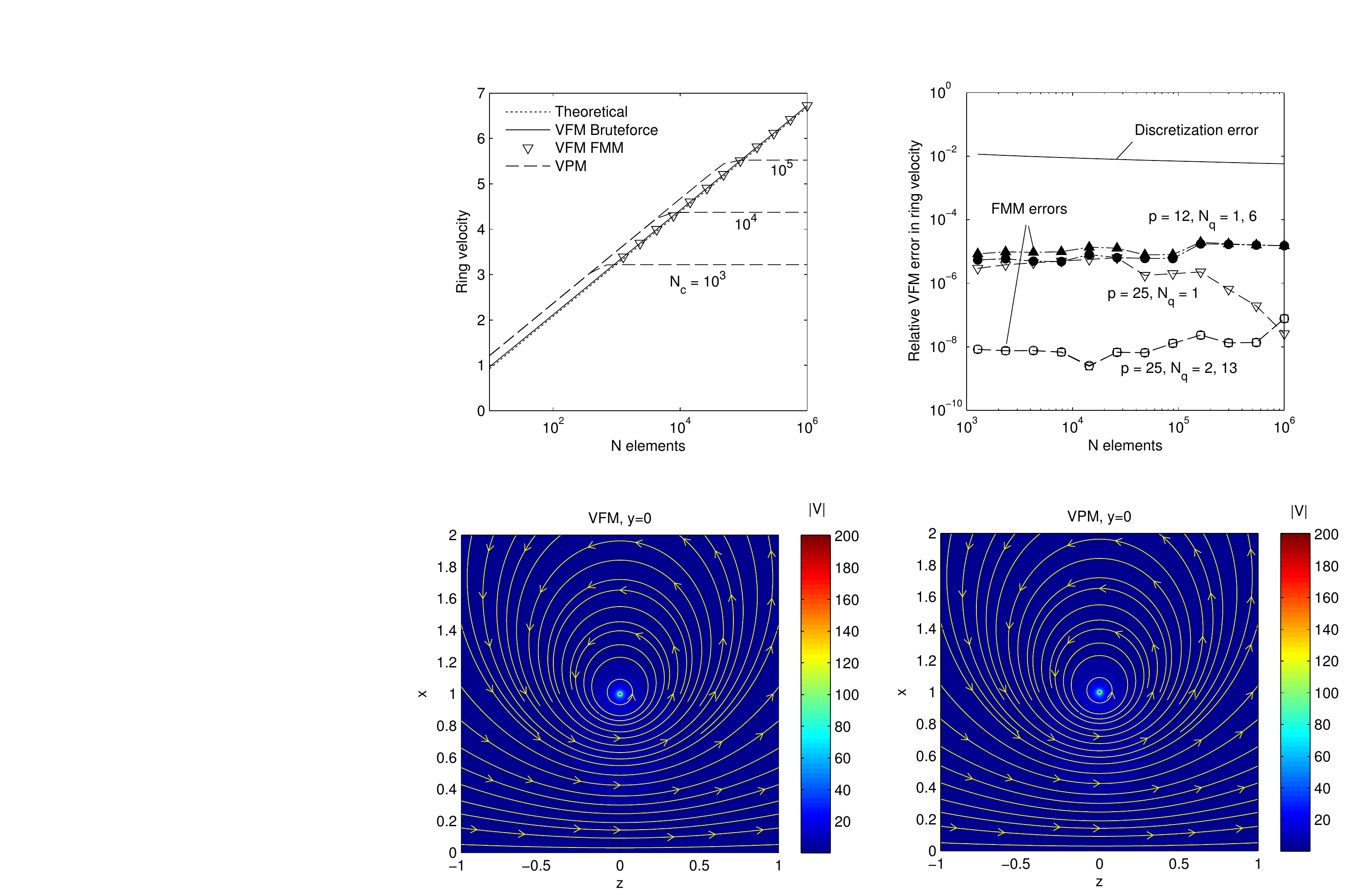}
\end{center}
\caption{The dependences of the vortex ring velocity and computational
errors on the number of discretization elements $N$ obtained by the VFM and
VPM (the top raw) and the respective velocity fields for $N=10^{6}$ (VPM)
and $N=490960$ (VFM) (the bottom raw) for $R=1$ and $\Gamma =2\protect\pi $.
The theoretical velocity is given by Eq. (\protect\ref{dis6}). The dashed
lines for the ring velocity correspond to the VPM computations with vortex
particle size parameter $a=2\protect\pi R/N_{c}$, where $N_{c}$ is shown
near the curves. The errors are plotted for different parameters values of $p
$ and $N_{q}$ controlling the accuracy of the FMM accelerated VFM. The
discretization error is due to the approximation of the circle by linear
elements and is not related to the FMM. }
\label{FigSingleRing}
\end{figure}

In Fig. \ref{FigSingleRing} we compared the velocity of the ring obtained
via VFM and VPM (average of all vortex element velocities) with theoretical
prediction (\ref{dis6}). The ring velocity found via the VFM using the brute
force method (no FMM) is about 1\% different from the value provided by Eq. (%
\ref{dis6}) and this difference decreases slowly with increasing $N$. So
this error is not related to the use of the FMM, which at $p=12$ and $p=25$
are several orders smaller, and can be referred to the discretization
scheme. Despite Eq. (\ref{vf7}) provides consistency of the quadrature and
FMM truncation errors it is sufficient to use $N_{q}=2$ and even $N_{q}=1$
at smaller $p$ or very large $N$. This can be explained by the fact that the
polygon approximation of the contour results in a globally non-smooth
integrand. In this case high order quadratures do not improve accuracy, but
the increase of the number of collocation points in a low order quadrature
does. In the VPM we specified the characteristic vortex blob size in Eq. (%
\ref{vem1}) as $a=2\pi R/N_{c}$. For $N\ll N_{c}$ the vortex blobs do not
overlap, while at $N\gg N_{c}$ the overlapping is high and we have a ring of
constant radius $\delta \sim a$, which does not depend on $N$ and so the
ring has a constant velocity (\ref{dis4}). We also computed the velocity
field using $N=10^{6}$, $N_{c}=5\cdot 10^{5}$ for the VPM and $N\approx
5\cdot 10^{5}$ for the VFM, which provides approximately the same ring
velocity. At $p=25$ the relative error between the numerical and analytical
solutions ($L_{\infty }$-norm) did not exceed $5\cdot 10^{-9}$ (eight
digits) for the distances larger than $2\delta $, where $\delta $ was found
from Eq. (\ref{dis4}). 

The brute force computations of the ring dynamics using the VFM or VPM are
stable, since all vortex elements are in symmetric positions. However, the
FMM errors introduce noise and asymmetry, which result in much faster
development of instabilities. So for ring dynamics computations at each time
step we applied a $O\left( N\right) $ stabilization procedure using a
low-pass FFT-based filtering of the curve shape.

\subsubsection{Pair of vortex rings}

Motion along the z-axis of two vortex rings of radii $R_{1}\left( t\right) $
and $R_{2}\left( t\right) $ and positions $Z_{1}\left( t\right) $ and $%
Z_{2}\left( t\right) $ can be found from the solution of the initial value
problem for the system of four ODEs 
\begin{equation}
\frac{dR_{j}}{dt}=V_{3-j}^{(r)}\left( R_{j},Z_{j};R_{3-j},Z_{3-j}\right)
,\quad \frac{dZ_{j}}{dt}=V_{j}^{(self)}\left( R_{j}\right)
+V_{3-j}^{(z)}\left( R_{j},Z_{j};R_{3-j},Z_{3-j}\right) ,\quad j=1,2,
\label{dis8}
\end{equation}
where $V_{j}^{(r)}$ and $V_{j}^{(z)}$ are components of the velocity field $%
\mathbf{V}_{j}\left( r,z;R_{j},Z_{j}\right) $ of the $j$th ring in
cylindrical coordinates $\left( r,z\right) $, and $V_{j}^{(self)}\left(
R_{j}\right) $ is the self-induced velocity, Eq. (\ref{dis4}). We integrated
Eq. (\ref{dis8}) using the 4th order Runge-Kutta method with controlled
relative error 10$^{-9}$.

\begin{figure}[tbh]
\begin{center}
\includegraphics[width=5.5in]
{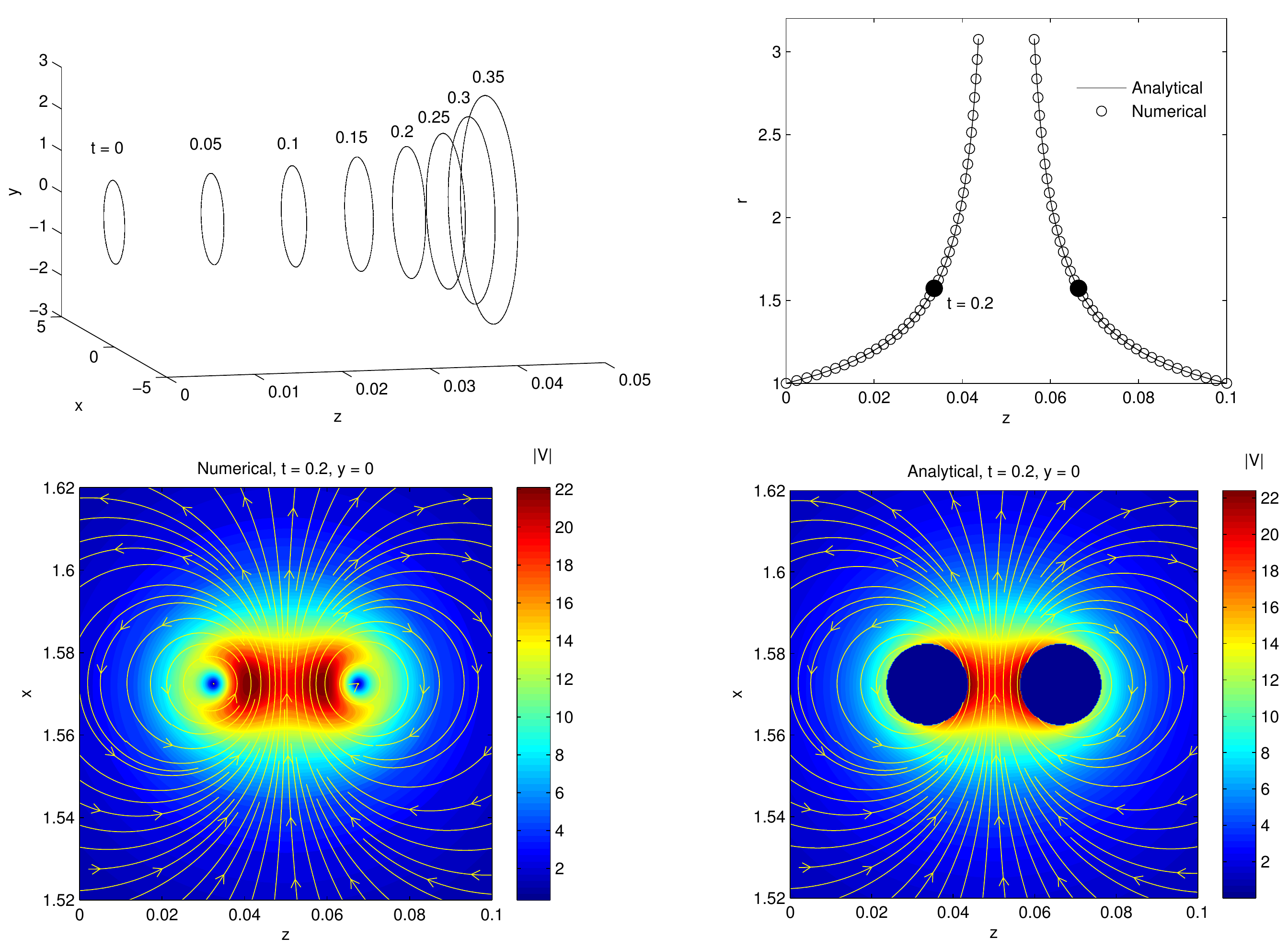}
\end{center}
\caption{A comparison of dynamics of colliding rings $\left( \Gamma
_{1}=-\Gamma _{2}=1\right) $, obtained by the VPM accelerated by the FMM ($%
p=25$) with analytical solution. Only ring 1 is shown in the upper left
picture. The velocity fields near the ring cores are shown for $t=0.2$. In
the plot of analytical solution the velocity field is zeroed within the core
radius $\protect\delta \approx 10^{-2}$ to show the radius.}
\label{FigCollidingRings}
\end{figure}

\begin{figure}[tbh]
\begin{center}
\includegraphics[width=5.5in, scale=1.35, trim=2in 0 0 1.5in]
{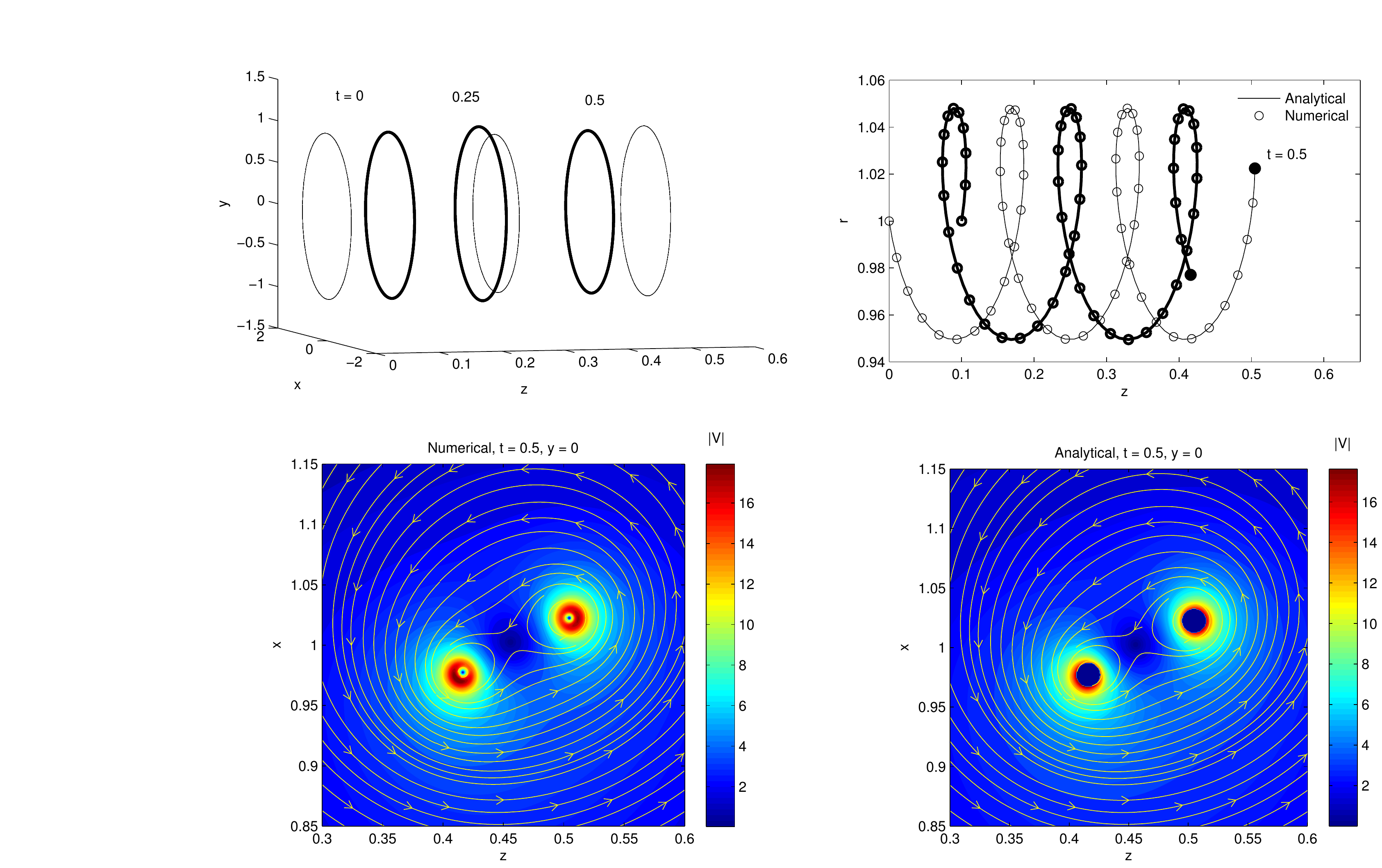}
\end{center}
\caption{The same as in Fig. \protect\ref{FigCollidingRings}, but for
leapfrogging rings $\left( \Gamma _{1}=\Gamma _{2}=1\right) $. In the upper
raw of pictures the data for ring \#1 are plotted by the thin lines and
markers, while the data for ring \#2 are plotted by the thick lines and
markers.}
\label{FigLeapfroggingRings}
\end{figure}

Numerical tests were performed for two cases with the same initial
conditions $R_{1}\left( 0\right) =R_{2}(0)=1,$ $Z_{1}\left( 0\right) =0,$ $%
Z_{2}\left( 0\right) =0.1$ and ring core radii $\delta _{1}=\delta
_{2}=\delta =9.966931\cdot 10^{-3}$, but different circulations $\Gamma
_{1}=-\Gamma _{2}=1$ and $\Gamma _{1}=\Gamma _{2}=1$. The first case
produces colliding rings, while the second case results in leapfrogging
rings, shown in Figs \ref{FigCollidingRings} and \ref{FigLeapfroggingRings}.
Vortex element simulations were performed using the VPM with $N=5\cdot
10^{4} $ elements per ring and blob size $a=2\pi /N_{c}$, $N_{c}=10^{3}$,
which determines the ring core radii provided above during all computed
times ($N_{c}\ll N$; for the colliding rings the initial radius increase was
about 3 times at $t=t_{\max }\sim 0.4$). The time integration in the VPM was
performed using a 4th order Runge-Kutta method with a constant time step $%
\Delta t=5\cdot 10^{-4}$ and an FFT-based low-pass shape filter (11 Fourier
modes retained for each ring). In all cases we used the FMM with $p=25$. All
errors were measured as the maximum relative errors, $\epsilon _{\infty }$ ($%
L_{\infty }$ norm). For the case of colliding rings positional (max of $Z^{%
\text{'}}$s and $R^{\text{'}}$s) error $\epsilon _{\infty }=$ $1.5\cdot
10^{-3}$ for $0\leqslant t\leqslant t_{\max }$, but $\epsilon _{\infty
}=1.4\cdot 10^{-5}$ for $0\leqslant t\leqslant t_{1}$ and $\epsilon _{\infty
}=1.2\cdot 10^{-6}$ for $0\leqslant t\leqslant t_{2}$, where at $t_{1}$ the
ring cores touch $\left( Z_{2}-Z_{1}=2\delta \right) $ and at $t=t_{2}$ we
have $Z_{2}-Z_{1}=4\delta $. For the case of leapfrogging rings positional
error $\epsilon _{\infty }<$ $6.2\cdot 10^{-8}$. This shows that the FMM
velocity computations are accurate and a larger error in the first case,
presumably, has a non-FMM related nature (due to the Gaussian spread of the
vortex blobs). Errors $\epsilon _{\infty }$ in the velocity fields in Figs %
\ref{FigCollidingRings} and \ref{FigLeapfroggingRings} at distances larger
than $2\delta $ were approximately the same for both cases, $8.9\cdot 10^{-5}
$ and $1.7\cdot 10^{-4}$, respectively. $\quad \ $ 

\bigskip

\section{Conclusion}

The main goal of this study was to develop an efficient method for fast
summation of elementary vortices. Numerical tests confirm the validity of
the theory presented and efficiency of the method. Our numerical results
show that one should expect an increase of the computation time by a factor
of approximately two for the velocity and vorticity stretching term
computations compared to a single harmonic function computations for the
same FMM octree and the truncation number. Compared to potential and
gradient computations for the scalar Laplace equation this increase varies
in the range from 1 to 2 times (average $\approx 1.4$) depending on
particular source/receiver distribution, truncation number, etc. These
numbers are in a good agreement with the theoretical FMM cost estimates.

An interesting observation from the study is that a general reconstruction
of vector fields from given curl and divergence can be obtained via the
present method, which operates only with two scalar potentials. This may
have application to many other fields of physics, including plasma physics,
electromagnetism, etc. In this sense the DCVLE appears to be a fundamental
equation, the solutions of which can be accelerated via the harmonic FMM. As
is shown, modifications of standard FMM programs are relatively easy, and
require tracking of two harmonic functions, and implementation of the
conversion operators used in each translation. Such operators are very
sparse and simple (especially for the case of the RCR-decomposition) and
their execution does not create substantial overheads. In terms of further
acceleration of computations it is natural to consider implementations of
the method on graphics processors (GPUs) for which the vortex methods are
developed and tested (e.g. \cite{Yokota2009:CPC}).

We also conducted some tests for the vortex filament and vortex particle
methods using the algorithm developed. These tests show a good accuracy of
computations compared to analytical solutions and indicate that the FMM
errors are typically much smaller than the errors of the VFM and VPM schemes
introduced by line integral discretization, vortex blob representation, time
integration errors, and interpolation procedures. The FMM brings the entire
VEM method to $O(N)$ complexity for highly clustered non-uniform data
distributions (e.g. vortex lines and sheets). Such acceleration is important
for numerous practical applications, e.g. for computation of vortical fields
generated by helicopters \cite{Hu2011:AHS}, while for such large scale
simulations further study is required.

\section{Acknowledgement}

Work partially supported by AFOSR under MURI Grant W911NF0410176 (PI
Dr.~J.~G. Leishman, monitor Dr.~D. Smith); and by Fantalgo, LLC.

\appendix

\section{Matrix representation of differential and conversion operators}

\subsection{Operator $\mathcal{D}_{\mathbf{r}}$}

To obtain the representation of this operator there is no need to consider
differential relations, since according to definition (\ref{bas2}) we have $%
\mathcal{D}_{\mathbf{r}}=r\left( \partial /\partial r\right) $ and from Eq. (%
\ref{rsbf2}) we have 
\begin{equation}
\mathcal{D}_{\mathbf{r}}R_{n}^{m}\left( \mathbf{r}\right) =nR_{n}^{m}\left( 
\mathbf{r}\right) ,\quad \mathcal{D}_{\mathbf{r}}S_{n}^{m}\left( \mathbf{r}%
\right) =-\left( n+1\right) S_{n}^{m}\left( \mathbf{r}\right) ,\quad
n=0,1,...  \label{dr1}
\end{equation}
This shows that matrices $\mathbf{D}_{\mathbf{r}}^{(R)}$ and $\mathbf{D}_{%
\mathbf{r}}^{(S)}$ representing this operator are diagonal and have entries 
\begin{equation}
\left( \mathbf{D}_{\mathbf{r}}^{(R)}\right) _{nn^{\prime }}^{mm^{\prime
}}=n\delta _{mm^{\prime }}\delta _{nn^{\prime }},\quad \left( \mathbf{D}_{%
\mathbf{r}}^{(S)}\right) _{nn^{\prime }}^{mm^{\prime }}=-\left( n+1\right)
\delta _{mm^{\prime }}\delta _{nn^{\prime }},\quad n=0,1,...  \label{dr2}
\end{equation}
where $\delta _{mm^{\prime }}$ is the Kronecker symbol.

Conversion operators (\ref{CVL14}) contain inverse operators $\mathcal{D}_{%
\mathbf{r}}^{-1}$ and $\left( \mathcal{I}+\mathcal{D}_{\mathbf{r}}\right)
^{-1}$, which also are diagonal. It may be a cause for concern for the
inverse operators that zeros appear on the diagonal of matrix $\mathbf{D}_{%
\mathbf{r}}^{(R)}$ and on the diagonal of matrix $\mathbf{I}+\mathbf{D}_{%
\mathbf{r}}^{(S)}$ at $n=0.$ However, these are easily dispensed with. For
the former case we note that harmonic $n=0$ corresponds to a constant basis
function $R_{0}^{0}\left( \mathbf{r}\right) .$ Eq. (\ref{CVL13}) shows that
this affects only the constant added to potential $\phi $, which obviously
does not affect the velocity field (\ref{4}), and can be set to an arbitrary
value, e.g. to zero. For the latter case, Eq. (\ref{CVL1}) shows that
harmonic $n=0$ in multipole expansion of function $\chi $ also does not
affect the velocity field, since 
\begin{equation}
\nabla \times \left( \mathbf{r}r^{-1}\right) =-\mathbf{r}\times \nabla
\left( r^{-1}\right) =r^{-2}\mathbf{r}\times \mathbf{r}=\mathbf{0.}
\label{dr3}
\end{equation}
Since operator $\left( \mathbf{I}+\mathbf{D}_{\mathbf{r}}^{(S)}\right) ^{-1}$
is needed only to determine converted function $\chi $ (see Eq. (\ref{CVL13}%
) this also can be set to zero. In other words, for the purpose of
computation of the conversion operators the inverse operators for singular
matrices can be defined as follows 
\begin{equation}
\left[ \left( \mathbf{D}_{\mathbf{r}}^{(R)}\right) ^{-1}\right] _{nn^{\prime
}}^{mm^{\prime }}=\delta _{mm^{\prime }}\delta _{nn^{\prime }}\left\{ 
\begin{array}{c}
n^{-1},\quad n>0 \\ 
0,\quad n=0%
\end{array}
\right. ,\quad \left[ \left( \mathbf{I}+\mathbf{D}_{\mathbf{r}}^{(S)}\right)
^{-1}\right] _{nn^{\prime }}^{mm^{\prime }}=\delta _{mm^{\prime }}\delta
_{nn^{\prime }}\left\{ 
\begin{array}{c}
-n^{-1},\quad n>0 \\ 
0,\quad n=0%
\end{array}
\right. .  \label{dr4}
\end{equation}

\subsection{Operator $\mathcal{D}_{\mathbf{t}}$}

From definitions (\ref{bas2}) and (\ref{mr5}) we have 
\begin{equation}
\mathcal{D}_{\mathbf{t}}=t_{x}\frac{\partial }{\partial x}+t_{y}\frac{%
\partial }{\partial y}+t_{z}\frac{\partial }{\partial z}=\frac{1}{2}\left[
\left( t_{x}-it_{y}\right) \mathcal{D}_{x+iy}+(t_{x}+it_{y})\mathcal{D}%
_{x-iy}\right] +t_{z}\mathcal{D}_{z}.  \label{dt1}
\end{equation}
Using Eq. (\ref{mr4}) we obtain 
\begin{eqnarray}
\mathcal{D}_{\mathbf{t}}R_{n}^{m} &=&\frac{1}{2}\left[ \left(
t_{y}+it_{x}\right) R_{n-1}^{m+1}-(t_{y}-it_{x})R_{n-1}^{m-1}\right]
-t_{z}R_{n-1}^{m},  \label{dt2} \\
\mathcal{D}_{\mathbf{t}}S_{n}^{m} &=&\frac{1}{2}\left[ \left(
t_{y}+it_{x}\right) S_{n+1}^{m+1}-(t_{y}-it_{x})S_{n+1}^{m-1}\right]
-t_{z}S_{n+1}^{m}.  \notag
\end{eqnarray}
Taking into account that matrices $\mathbf{D}_{\mathbf{t}}^{(R)}$ and $%
\mathbf{D}_{\mathbf{t}}^{(S)}$ are transposed to the reexpansion matrices,
we obtain 
\begin{eqnarray}
\left( \mathbf{D}_{\mathbf{t}}^{(R)}\right) _{nn^{\prime }}^{mm^{\prime }}
&=&\delta _{n+1,n^{\prime }}\left[ \frac{1}{2}\left( t_{y}+it_{x}\right)
\delta _{m-1,m^{\prime }}-\frac{1}{2}(t_{y}-it_{x})\delta _{m+1,m^{\prime
}}-t_{z}\delta _{mm^{\prime }}\right] ,  \label{dt3} \\
\left( \mathbf{D}_{\mathbf{t}}^{(S)}\right) _{nn^{\prime }}^{mm^{\prime }}
&=&\delta _{n-1,n^{\prime }}\left[ \frac{1}{2}\left( t_{y}+it_{x}\right)
\delta _{m-1,m^{\prime }}-\frac{1}{2}(t_{y}-it_{x})\delta _{m+1,m^{\prime
}}-t_{z}\delta _{mm^{\prime }}\right] .  \notag
\end{eqnarray}

\subsection{Operator $\mathcal{D}_{\mathbf{r\times t}}$}

From definitions (\ref{bas2}) and (\ref{mr5}) we have 
\begin{eqnarray}
\mathcal{D}_{\mathbf{r\times t}} &=&\left( yt_{z}-zt_{y}\right) \frac{%
\partial }{\partial x}+\left( zt_{x}-xt_{z}\right) \frac{\partial }{\partial
y}+\left( xt_{y}-yt_{x}\right) \frac{\partial }{\partial z}  \label{drt1} \\
&=&-i(t_{x}+it_{y})\left[ \xi _{-}\mathcal{D}_{z}-\frac{1}{2}z\mathcal{D}%
_{x-iy}\right] +i(t_{x}-it_{y})\left[ \xi _{+}\mathcal{D}_{z}-\frac{1}{2}z%
\mathcal{D}_{x+iy}\right] +it_{z}\left[ \xi _{-}\mathcal{D}_{x+iy}-\xi _{+}%
\mathcal{D}_{x-iy}\right] ,  \notag \\
\quad \xi _{\pm }&=&\frac{x\pm iy}{2}.  \notag
\end{eqnarray}
Consider first action of this operator on basis functions $R_{n}^{m}\left( 
\mathbf{r}\right) $. The following relations derived in \cite%
{Gumerov2006:JCP} are useful in this case: 
\begin{eqnarray}
\xi _{+}R_{n}^{m} &=&-i\frac{n+m+2}{2}R_{n+1}^{m+1}-\frac{i}{2}zR_{n}^{m+1},
\label{drt2} \\
\xi _{-}R_{n}^{m} &=&-i\frac{n-m+2}{2}R_{n+1}^{m-1}-\frac{i}{2}zR_{n}^{m-1}.
\notag
\end{eqnarray}
We have then, using these relations and Eq. (\ref{mr4}) 
\begin{eqnarray}
\left[ \xi _{-}\mathcal{D}_{z}-\frac{1}{2}z\mathcal{D}_{x-iy}\right]
R_{n}^{m} &=&-\xi _{-}R_{n-1}^{m}-\frac{1}{2}izR_{n-1}^{m-1}=i\frac{n-m+1}{2}%
R_{n}^{m-1},  \label{drt3} \\
\left[ \xi _{+}\mathcal{D}_{z}-\frac{1}{2}z\mathcal{D}_{x+iy}\right]
R_{n}^{m} &=&-\xi _{+}R_{n-1}^{m}-\frac{1}{2}izR_{n-1}^{m+1}=i\frac{n+m+1}{2}%
R_{n}^{m+1},  \notag \\
\left[ \xi _{-}\mathcal{D}_{x+iy}-\xi _{+}\mathcal{D}_{x-iy}\right]
R_{n}^{m} &=&i\xi _{-}R_{n-1}^{m+1}-i\xi _{+}R_{n-1}^{m-1}=-mR_{n}^{m}. 
\notag
\end{eqnarray}
Now we obtain from Eq. (\ref{drt1}) 
\begin{equation}
\mathcal{D}_{\mathbf{r\times t}}R_{n}^{m}=(t_{x}+it_{y})\frac{n-m+1}{2}%
R_{n}^{m-1}-(t_{x}-it_{y})\frac{n+m+1}{2}R_{n}^{m+1}-it_{z}mR_{n}^{m}.
\label{drt4}
\end{equation}
To get a similar relation for basis functions $S_{n}^{m}\left( \mathbf{r}%
\right) $ we can use relation (\ref{rsbf4}). Using identity $\left( \mathbf{%
r\times t}\right) \cdot \nabla \left[ f\left( r\right) g\left( \mathbf{r}%
\right) \right] =f\left( r\right) \left( \mathbf{r\times t}\right) \cdot
\nabla g\left( \mathbf{r}\right) $ and Eq. (\ref{drt4}), we obtain 
\begin{eqnarray}
\mathcal{D}_{\mathbf{r\times t}}S_{n}^{m} &=&\left( -1\right) ^{n+m}\left(
n-m\right) !(n+m)!r^{-2n-1}\mathcal{D}_{\mathbf{r\times t}}R_{n}^{m}
\label{drt5} \\
&=&-(t_{x}+it_{y})\frac{n+m}{2}S_{n}^{m-1}+(t_{x}-it_{y})\frac{n-m}{2}%
S_{n}^{m+1}-it_{z}mS_{n}^{m}.  \notag
\end{eqnarray}

Expressions for the representing matrices for the local and multipole bases
follow from Eqs (\ref{drt4}) and (\ref{drt5}): 
\begin{eqnarray}
\left( \mathbf{D}_{\mathbf{r}\times \mathbf{t}}^{(R)}\right) _{nn^{\prime
}}^{mm^{\prime }} &=&\delta _{nn^{\prime }}\left[ (t_{x}+it_{y})\frac{%
n^{\prime }-m^{\prime }+1}{2}\delta _{m+1,m^{\prime }}-(t_{x}-it_{y})\frac{%
n^{\prime }+m^{\prime }+1}{2}\delta _{m-1,m^{\prime }}-it_{z}m^{\prime
}\delta _{mm^{\prime }}\right] ,  \notag \\
\left( \mathbf{D}_{\mathbf{r}\times \mathbf{t}}^{(S)}\right) _{nn^{\prime
}}^{mm^{\prime }} &=&\delta _{nn^{\prime }}\left[ -(t_{x}+it_{y})\frac{%
n^{\prime }+m^{\prime }}{2}\delta _{m+1,m^{\prime }}+(t_{x}-it_{y})\frac{%
n^{\prime }-m^{\prime }}{2}\delta _{m-1,m^{\prime }}-it_{z}m^{\prime }\delta
_{mm^{\prime }}\right] .  \label{drt7}
\end{eqnarray}

\subsection{Conversion operators}

It is not difficult to obtain matrix representations for the conversion
operator from Eqs (\ref{CVL4}) and (\ref{CVL14}) and expressions for the
differential operators derived above. A more compact form relating the
expansion coefficients of functions in Eq. (\ref{CVL13}) for the $R$
expansions is 
\begin{eqnarray}
\widetilde{\phi }_{n}^{m} &=&\widehat{\phi }_{n}^{m}-\frac{1}{n}\left[
(t_{x}+it_{y})\frac{n-m}{2}\widehat{\chi }_{n}^{m+1}-(t_{x}-it_{y})\frac{n+m%
}{2}\widehat{\chi }_{n}^{m-1}-it_{z}m\widehat{\chi }_{n}^{m}\right] ,\quad
\left( \widetilde{\phi }_{0}^{0}=\widehat{\phi }_{0}^{0}\right) 
\label{drt10} \\
\widetilde{\chi }_{n}^{m} &=&\widehat{\chi }_{n}^{m}+\frac{1}{n+1}\left[ 
\frac{1}{2}(t_{y}+it_{x})\widehat{\chi }_{n+1}^{m-1}-\frac{1}{2}%
(t_{y}-it_{x})\widehat{\chi }_{n+1}^{m+1}-t_{z}\widehat{\chi }_{n+1}^{m}%
\right] .  \notag
\end{eqnarray}%
Similarly, for the $S$ expansions we have 
\begin{eqnarray}
\widetilde{\phi }_{n}^{m} &=&\widehat{\phi }_{n}^{m}+\frac{1}{n+1}\left[
-(t_{x}+it_{y})\frac{n+m+1}{2}\widehat{\chi }_{n}^{m+1}+(t_{x}-it_{y})\frac{%
n-m+1}{2}\widehat{\chi }_{n}^{m-1}-it_{z}m\widehat{\chi }_{n}^{m}\right] ,
\label{drt11} \\
\widetilde{\chi }_{n}^{m} &=&\widehat{\chi }_{n}^{m}-\frac{1}{n}\left[ \frac{%
1}{2}(t_{y}+it_{x})\widehat{\chi }_{n-1}^{m-1}-\frac{1}{2}(t_{y}-it_{x})%
\widehat{\chi }_{n-1}^{m+1}-t_{z}\widehat{\chi }_{n-1}^{m}\right] ,\quad
\left( \widetilde{\chi }_{0}^{0}=\widehat{\chi }_{0}^{0}\right) .  \notag
\end{eqnarray}

\section{Some line integrals}

\subsection{Velocity field of linear element}

Consider velocity field, Eq. (\ref{dis1}), of a linear vortex element of
constant circulation $\Gamma $ with end points $\mathbf{x}^{(1)}$ and $%
\mathbf{x}^{(2)}$. This integral diverges for $\mathbf{y}\in C$. For $%
\mathbf{y}\notin C$ we change the integration variable as 
\begin{equation}
\mathbf{x}=\mathbf{x}^{(c)}+\frac{1}{2}\mathbf{l}\xi ,\quad \mathbf{x}^{(c)}=%
\frac{1}{2}\left( \mathbf{x}^{(2)}+\mathbf{x}^{(1)}\right) ,\quad \mathbf{l}=%
\mathbf{x}^{(2)}-\mathbf{x}^{(1)},\quad d\mathbf{l}\left( \mathbf{x}\right) =%
\frac{1}{2}\mathbf{l}d\xi ,  \label{vf2}
\end{equation}
to obtain 
\begin{equation}
\mathbf{v}\left( \mathbf{y}\right) =\frac{\Gamma }{8\pi }\mathbf{l}\times
\left( \mathbf{y-x}^{(c)}\right) \int_{-1}^{1}\frac{d\xi }{\left| \mathbf{y-x%
}^{(c)}+\frac{1}{2}\mathbf{l}\xi \right| ^{3}}.  \label{vf3}
\end{equation}
The definite integral here can be computed using the primitive (can be
checked by differentiation) 
\begin{equation}
\int \frac{d\xi }{\left| \mathbf{a}+\mathbf{b}\xi \right| ^{3}}=\frac{%
b^{2}\xi +\left( \mathbf{a}\cdot \mathbf{b}\right) }{\left[
a^{2}b^{2}-\left( \mathbf{a}\cdot \mathbf{b}\right) ^{2}\right] \left| 
\mathbf{a}+\mathbf{b}\xi \right| }+C,\quad a=\left| \mathbf{a}\right| ,\quad
b=\left| \mathbf{b}\right| .  \label{vf4}
\end{equation}
Further we note that all vectors in the resulting formula according to Eq. (%
\ref{vf2}) can be expressed in terms of $\mathbf{r}^{(1)}=\mathbf{y-x}^{(1)}$
and $\mathbf{r}^{(2)}=\mathbf{y-x}^{(2)}$, which results in expression (\ref%
{dis2}).

\subsection{Far field expansions}

Far field expansion of the integrand for elementary velocity field in Eq. (%
\ref{vf3}) can be obtained similarly to derivation of Eq. (\ref{eve4}) from
Eq. (\ref{eve1}). Indeed, we have for expansion center $\mathbf{x}_{\ast }$, 
\begin{eqnarray}
\mathbf{v}\left( \mathbf{y}\right)  &=&\int_{-1}^{1}\nabla \times \frac{%
\Gamma }{8\pi }\frac{\mathbf{l}d\xi }{\left\vert \mathbf{y-x}^{(c)}+\frac{1}{%
2}\mathbf{l}\xi \right\vert }=\sum_{n=0}^{\infty }\sum_{m=-n}^{n}\widetilde{g%
}_{n}^{m}\mathbf{F}_{n}^{m}\left( \mathbf{r}\right) ,  \label{vf5} \\
\widetilde{g}_{n}^{m} &=&\int_{-1}^{1}R_{n}^{-m}(-\mathbf{r}^{(c)}+\frac{1}{2%
}\mathbf{l}\xi )d\xi ,\quad \mathbf{r}=\mathbf{y-x}_{\ast },\quad \mathbf{r}%
^{(c)}=\mathbf{x}^{(c)}-\mathbf{x}_{\ast },  \notag
\end{eqnarray}%
Hence, Eqs (\ref{eve2}) and (\ref{eve3}) can be used to expand potentials $%
\phi $ and $\chi $ into the harmonic series. Coefficients of these series
are provided by Eq. (\ref{eve4}), where one simply should use $\frac{\Gamma 
}{8\pi }\mathbf{l}$ instead of vector $\mathbf{\omega }_{l}$ and
coefficients $\widetilde{g}_{n}^{m}$ instead of $g_{nl}^{m}$. The problem
then is to compute $\widetilde{g}_{n}^{m}$ in Eq. (\ref{vf5}). These
coefficients can be computed in a very straightforward way using
Gauss-Legendre quadrature with weights $w_{j}$ and abscissas $\xi _{j}$ (see 
\cite{Abramowitz1972}), 
\begin{equation}
\widetilde{g}_{n}^{m}=\sum_{j=1}^{N_{q}}w_{j}R_{n}^{-m}(-\mathbf{r}^{(c)}+%
\frac{1}{2}\mathbf{l}\xi _{j}).  \label{vf6}
\end{equation}%
Note now that functions $R_{n}^{-m}(-\mathbf{r}^{(c)}+\frac{1}{2}\mathbf{l}%
\xi )$ are polynomials of degree $n$ of $\xi $ (see \cite{Gumerov2006:JCP}).
Hence, the Gauss-Legendre quadrature (\ref{vf6}) provides an exact result
for $N_{q}>n/2$. Furthermore, for application of the FMM we truncate all
series with $p$ terms to provide a required accuracy. Therefore, the range
of $n$ needed is limited as $n\leqslant p-1$, and choice

\begin{equation}
N_{q}=\left[ \frac{p-1}{2}\right] +1,  \label{vf7}
\end{equation}
provides an exact result for all harmonics needed for application of the FMM.

\subsection{Velocity of the vortex ring without self-induction of small
elements}

We exclude an $\epsilon $ vicinity of a point \textbf{y} on the ring by
putting its self-induction to zero. In this case we have 
\begin{equation}
\mathbf{v}\left( \mathbf{y}\right) =\frac{\Gamma }{4\pi }\int_{C\backslash
C_{\epsilon }}\frac{d\mathbf{l}\left( \mathbf{x}\right) \times \left( 
\mathbf{y-x}\right) }{\left\vert \mathbf{y-x}\right\vert ^{3}}.  \label{vf8}
\end{equation}%
Now for a ring in $z$ plane and $\mathbf{y}$ located on the $x$ axis we can
express Cartesian coordinates of vectors in Eq. (\ref{vf8}) as 
\begin{equation}
\mathbf{x=}R\left( \cos \varphi ,\sin \varphi ,0\right) ,\quad \mathbf{y=}%
R\left( 1,0,0\right) ,\quad d\mathbf{l}\left( \mathbf{x}\right) =\mathbf{i}%
_{\varphi }Rd\varphi =\left( -\sin \varphi ,\cos \varphi ,0\right) Rd\varphi
,  \label{vf9}
\end{equation}%
where $\varphi $ is the polar angle for the reference frame centered at the
ring center. Substituting this into Eq. (\ref{vf8}), we obtain 
\begin{eqnarray}
\mathbf{v}\left( \mathbf{y}\right)  &=&\frac{\Gamma }{4\pi R\cdot 2^{3/2}}%
\mathbf{i}_{z}\int_{\epsilon /2}^{2\pi -\epsilon /2}\frac{1}{\left( 1-\cos
\varphi \right) ^{1/2}}d\varphi =\frac{\Gamma }{4\pi R}\mathbf{i}%
_{z}\int_{\epsilon /4}^{\pi /2}\frac{d\psi }{\sin \psi }  \label{vf10} \\
&=&-\frac{\Gamma }{4\pi R}\mathbf{i}_{z}\ln \left\vert \tan \frac{\epsilon }{%
8}\right\vert =\frac{\Gamma }{4\pi R}\mathbf{i}_{z}\left[ \ln \frac{8}{%
\epsilon }+O\left( \epsilon ^{2}\right) \right] .  \notag
\end{eqnarray}%
If the size of the line element in the VFM is $l$, then $\epsilon R\approx 2l
$ and the total number of elements is $N=2\pi R/l$. Hence 
\begin{equation}
V_{\epsilon }^{(self)}=\frac{\Gamma }{4\pi R}\left[ \ln \frac{4R}{l}+O\left( 
\frac{l}{R}\right) ^{2}\right] =\frac{\Gamma }{4\pi R}\left( \ln \frac{2N}{%
\pi }+O\left( \frac{1}{N^{2}}\right) \right) .  \label{vf11}
\end{equation}%
We also can determine the \textquotedblleft effective\textquotedblright\
radius of the core for a given discretization (compare Eqs (\ref{dis4}) and (%
\ref{vf11})), 
\begin{equation}
\delta \sim 2le^{-1/4}\approx 1.5576l,\quad \frac{\delta }{R}\sim \frac{4\pi
e^{-1/4}}{N}\approx \frac{9.7867}{N}.  \label{vf12}
\end{equation}

\end{document}